\documentclass[%
 aip,
 amsmath,amssymb,
 reprint,%
]{revtex4-1}

\usepackage{graphicx}
\usepackage{dcolumn}
\usepackage{bm}
\usepackage{float}

\usepackage[utf8]{inputenc}
\usepackage[T1]{fontenc}
\usepackage{mathptmx}
\usepackage{etoolbox}
\usepackage[normalem]{ulem}

\usepackage{color}

\newcommand{\ks}{\textcolor{black}}

\makeatletter
\def\@email#1#2{%
 \endgroup
 \patchcmd{\titleblock@produce}
  {\frontmatter@RRAPformat}
  {\frontmatter@RRAPformat{\produce@RRAP{*#1\href{mailto:#2}{#2}}}\frontmatter@RRAPformat}
  {}{}
}%
\makeatother
\begin{document}

\preprint{AIP/123-QED}

\title{Physics of drying complex fluid drop: flow field, pattern formation, and desiccation cracks}
\author{Ranajit Mondal}
 \affiliation{Department of Chemical Engineering, Indian Institute of Technology Hyderabad, Kandi - 502 284, Telangana, India}
\author{Hisay Lama}
 \affiliation{Department of Physics, The University of Tokyo, Tokyo 113-8654, Japan}
 \author{Kirti Chandra Sahu}
 \affiliation{Department of Chemical Engineering, Indian Institute of Technology Hyderabad, Kandi - 502 284, Telangana, India}
\email{ranajit@che.iith.ac.in; hisaylama@noneq.phys.s.u-tokyo.ac.jp; ksahu@che.iith.ac.in}

\date{\today}

\begin{abstract}
Drying complex fluids is a common phenomenon where a liquid phase transforms into a dense or porous solid. This transformation involves several physical processes, such as the diffusion of liquid molecules into the surrounding atmosphere and the movement of dispersed phases through evaporation-driven flow. As a result, the solute forming a dried deposit exhibits unique patterns and often displays structural defects like desiccation cracks, buckling, or wrinkling. Various drying configurations have been utilized to study the drying of colloids, the process of their consolidation, and fluid-flow dynamics. This review focuses on the drying of colloids and the related phenomena, specifically the drying-induced effects observed during sessile drop drying. We first present a theoretical overview of the physics of drying pure and binary liquid droplets, followed by drying colloidal droplets. Then, we explain the phenomena of pattern formation and desiccation cracks. Additionally, the article briefly describes the impact of evaporation-driven flows on the accumulation of particles and various physical parameters that influence deposit patterns and cracks.
\end{abstract}

\maketitle

\section{Introduction}
\label{sec:intro}
Drying complex fluids, viz. particulate dispersion, the polymer solution, or biological fluids, is a fascinating phenomenon where a liquid phase is transformed into a dense or porous solid and has technological importance~\cite{giorgiutti2018drying}. The drying methods are frequently used in materials science to produce high-performance coatings on solid surfaces for technologies related to paints, microelectronics, and cosmetics. In all these aforementioned applications, the method of creating films from colloidal dispersion is becoming increasingly significant. A colloidal dispersion is a heterogeneous system in which particles known as the dispersed phase are present in a medium (solid, liquid, or gas), often referred to as the continuous phase. The typical dimension of the dispersed phase ranges from 1\,nm to 1\,$\mu$m. The dispersed particles in the medium experience several body forces and interaction forces, such as hydrodynamic, gravitational, Brownian, van der Waals, electrostatic and capillary forces~\cite{russel1991colloidal, mewis2012colloidal}. These body forces are exerted on the individual particles, while the interaction forces act between the multiple particles. The interaction forces acting between these particles can either be attractive or repulsive in nature. For example, van der Waals and capillary forces are attractive, while the electrostatic forces are repulsive. The relative strength of these interaction forces (van der Waals and electrostatic) determines the stability of a colloidal  dispersion~\cite{mewis2012colloidal}.

Colloidal dispersion as a model system has been extensively used in various technological domains~\cite{russel1991colloidal}. Drying of colloid is a ubiquitous phenomenon we see in our daily life~\cite{routh2013drying, goehring2015desiccation}. A typical example is a coffee stain mark that forms after complete evaporation of fluid from a drop of the coffee~\cite{deegan1997capillary,deegan2000contact}. Apart from drying of coffee, a few other examples are - drying ink~\cite{park2006control}, drying of paint~\cite{abas2003classification,giorgiutti2016painting} and drying mud~\cite{goehring2009nonequilibrium, goehring2013evolving} etc. In principle, drying involves an accumulation of particles, their rearrangement, and the evaporation of solvent liquid. The development of colloids into a dried particle deposit is commonly believed to be dictated by the advection and diffusion processes during drying. The particle deposit formed via evaporation-driven processes results in many patterns, accompanying structural defects such as desiccation cracks~\cite{routh2013drying}. In addition to purely scientific curiosity, understanding the mechanisms underlying this phenomenon is essential for various technological applications, such as fabricating DNA/RNA micro-arrays, controlled evaporative self-assembly (CESA), and more traditional processes like ink-jet printing and coating. The evaporation of colloidal droplet can also be used to fabricate self-assembled microstructures that results in an ordered crystal~\cite{juillerat2006formation, zhang2009self}. This ordered structure has potential application in the fabrication of optical devices. Thus, evaporation and subsequent deposition of dispersed particles on the substrate have been investigated for both \ks{single-component (pure)\cite{deegan1997capillary,sefiane2014patterns} and multi-component liquid droplets. One example of the latter is the ethanol-water mixture, which is a binary fluid and has been discussed in later sections \cite{katre2021evaporation,katre2022experimental,katre2020evaporation}.}

Another active area of research is the nucleation of defects like desiccation cracks in particulate film and how they self-organize into different morphologies.~\cite{goehring2015desiccation}. The subject of desiccation cracks has received good attention for the past few decades and has become a popular subject of research among the diverse community of researchers. There are several research domains wherein the desiccation cracks are prevalent, and they are - geosciences~\cite{muller1998experimental}, crackle-lithography~\cite{nam2012patterning, han2013drying, kim2015cracking}, coating technology~\cite{routh2013drying}, paint restoration~\cite{giorgiutti2016painting} and diagnostic~\cite{brutin2012influence}. Note that the formation of cracks in the deposit is undesirable for most practical applications, but the possibility of tailoring them into a regular pattern has potential applications in lithography~\cite{nam2012patterning, han2013drying}. This has further motivated the researchers to manoeuvre multiple techniques for tailoring the cracks into the desired pattern.

\begin{figure}[ht]
    \centering
    \includegraphics[width = \linewidth]{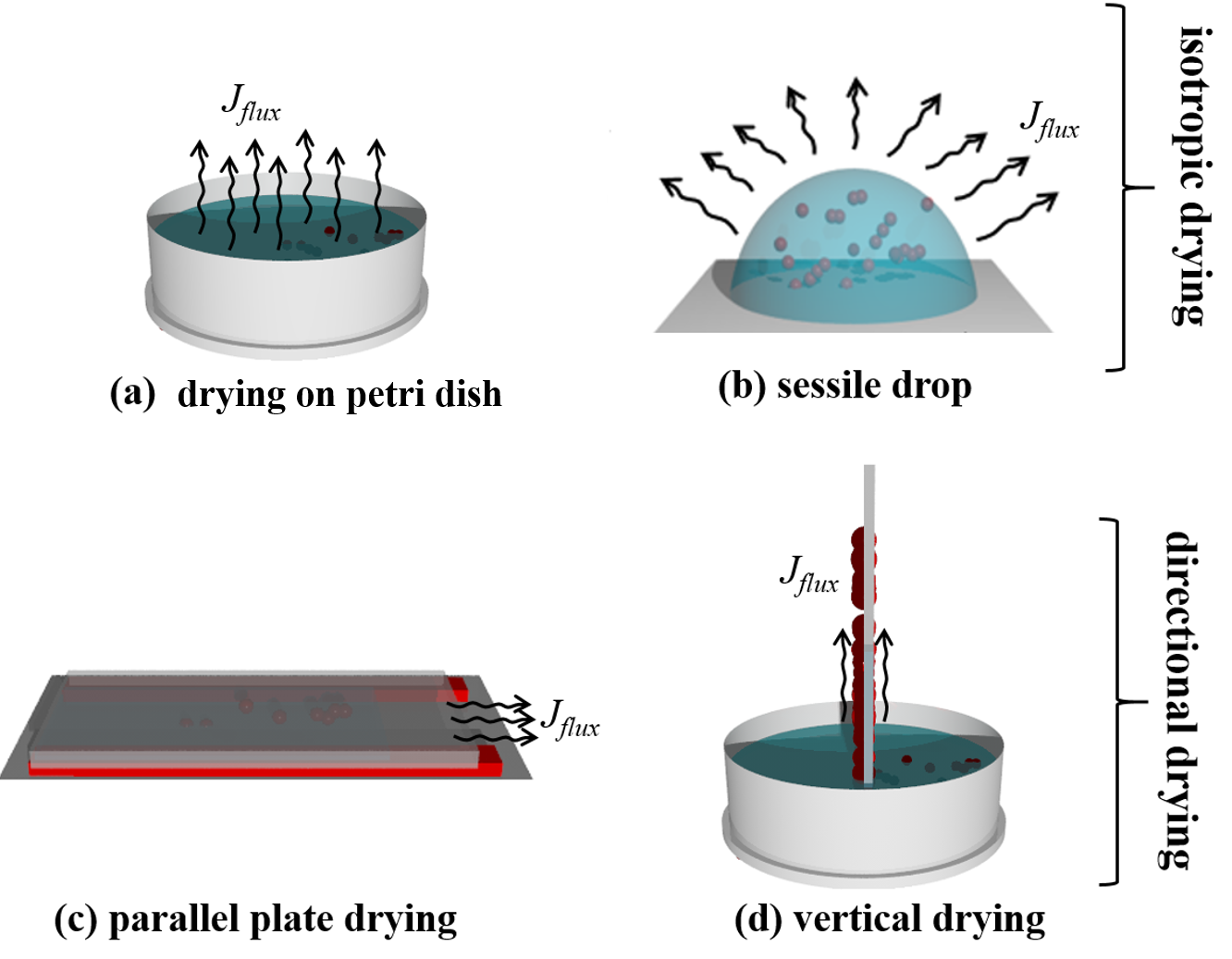}
    \caption{Schematic of various drying configurations, namely, isotropic and directional. Here, $J_{flux}$ denotes the evaporative flux. The drying of colloidal dispersion (a) on a  Petri dish, (b) in sessile configuration, (c) inside the parallel plate and (d) via vertical drying configuration. The arrows indicate the direction of diffusion of vapor to an ambient atmosphere. \textit{This figure is taken from PhD thesis of Hisay Lama at IIT Madras}.}
    \label{fig_1}
\end{figure}

In the laboratory, the physics of drying and the pattern formation is commonly studied by conducting the model evaporation experiments~\cite{bacchin2018drying}, namely, isotropic or directional drying as shown in Fig.~\ref{fig_1} (a)-(d). Drying colloids on Petri dish or microscope glass slides can be classified as isotropic drying while drying inside a parallel plate or via vertical deposition can be classified as directional drying. Conducting experiments in these configurations allows one to theoretically model fluid flow dynamics and quantify physical parameters, such as diffusion coefficient and evaporative flux associated with the drying process. The sessile drop configuration has gotten more attention due to its ubiquity among these drying configurations. The sessile configuration has an axisymmetric geometry, and quantities that characterize such drying phenomena can easily be extracted in 3D-cylindrical polar coordinates. Drying in sessile configuration has been widely exploited to understand the physics of drying and investigate the drying-related phenomenon. Therefore, we will describe the drying-related phenomenon primarily for sessile drop configuration. In this article, we first review the phenomena of drying pure and binary liquid droplets in a model sessile drop configuration. Then we describe the drying of a particulate droplet and two important attributes of it, namely (a) pattern formation and (b) desiccation cracks. 

\section{Evaporation of pure and binary drops}
An evaporating sessile droplet undergoes intricate and interconnected physical processes, which are associated with the mass flux arising due to (i) diffusion of vapour in the ambient air, (ii) free convection and (iii) natural convection of air from the hot substrate to the cool ambient. In the following, we discuss about each effect separately, then we combine them to get the overall evaporation dynamics of pure and binary liquid drops.    

\subsection{Pure liquid droplets}

\subsubsection{Diffusion model}
\begin{figure}
\centering
\includegraphics[width=0.4\textwidth]{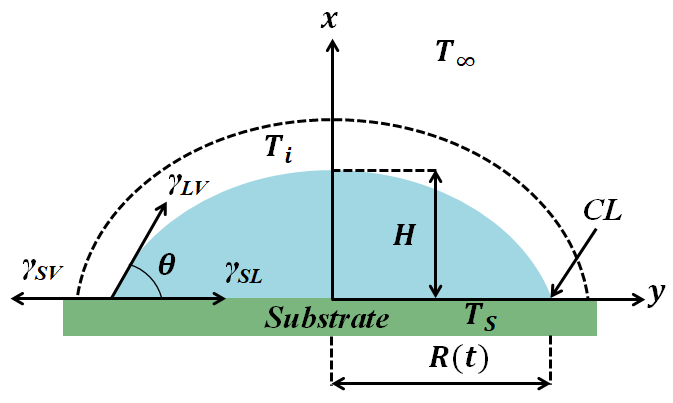} 
\caption{Schematic diagram of a sessile droplet on a substrate. Here, $T_s$, $T_i$ and $T_\infty$ denote the substrate temperature, temperature at the liquid-vapour interface and temperature of the ambient far away from the droplet, respectively; $\theta$, $H$ and $R(t)$ represent the contact angle, height and wetting radius of the droplet, respectively.}
\label{fig:sch}
\end{figure}

A droplet placed on a solid substrate exhibits a spherical cap geometry and forms a three-phase contact line at the edge where the solid, liquid and gas co-exist (depicted in Fig.~\ref{fig:sch}). This configuration is popularly known as the sessile drop. In principle, the process of drying is characterized by two-time scales viz. diffusion time scale ($\tau_{diff}$) and evaporation time scale ($\tau_{evap}$). The former is related to the diffusion of vapour into the ambient atmosphere, while the latter is the time required to evaporate completely, which are given by~\cite{giorgiutti2018drying}
\begin{eqnarray}
    \tau_{diff} \sim \frac {R^{2}}{\cal D}, \label{eq1} ~ {\rm and} \\
    \tau_{evap} \sim \frac {R}{{dV}/dt}, \label{eq2}
\end{eqnarray}
where $\cal D$ is the diffusion coefficient of vapour into the atmosphere, $R$ is the radius of the drop (indicated in Fig.~\ref{fig:sch}) and $dV/dt$ is the rate of change in volume of the drop. The relative magnitude of these timescales dictates the evaporation mechanism. For example, when $\tau_{diff}/\tau_{evap} \gg 1$, the evaporation is called diffusion limited. 

The diffusion-based vapour transport mechanism dominates the evaporation process at room temperature \citep{brutin2015droplet}. Assuming that a sessile droplet exhibits a spherical cap profile, the volume of the droplet can be calculated as 
\begin{equation}
V (t) = {\pi R^3 \over 3} {(1-\cos \theta)^2 (2 + \cos \theta) \over \sin^3 \theta}.
\end{equation}
A schematic diagram of a sessile droplet is shown in Fig. \ref{fig:sch}.
Further, it is assumed that the droplet evaporation occurs in an isothermal condition and the contact angle, $\theta<90^\circ$. The vapour concentration at the liquid-vapour interface is assumed to be at the saturated condition, $c_{sat}(T_s)$ and the vapour concentration in the region far away from the droplet $c_\infty(T_\infty)$ is ${\cal H} c_{sat}(T_s)$. Here, ${\cal H}$ is the relative humidity of ambient air. Under these conditions, mass evaporation rate due to diffusion $(\left(\frac{dm}{dt}\right)_d)$ for a pure droplet is given by \citep{sobac2012,carle2016}
\begin{equation}
\left(\frac{dm}{dt}\right)_d =  \pi R {\cal D} {\cal M} \left[c_{sat} (T_s) - c_\infty (T_\infty)\right] f(\theta), \label{case1a}
\end{equation}
where ${\cal M}$ denotes the molecular weight of the liquid. Here, $f(\theta) = 1.3 + 0.27 \theta^2$, as given in \citet{hu2002}.

At a given instant, the local evaporative flux on the surface of a drop, $J_{flux}$, is a function of the radial location, $r$. This is given by~\cite{deegan1997capillary,deegan2000contact}
\begin{equation}
J_{flux} (r) \propto \Big(1 - \frac{r}{R}\Big)^{-\big(\frac {\pi - 2\theta}{2\pi - 2\theta}\big)}.
\label{eq4}
\end{equation}
Eq. (\ref{eq4}) can be further simplified as
\begin{equation}
    J_{flux} (r) = J_0 \Big(1 - \frac{r}{R}\Big)^{-\big(\frac {\pi - 2\theta}{2\pi - 2\theta}\big)}, 
\label{eq5}
\end{equation}
where $J_0$ is a proportionality constant. As evident from Eqs.~(\ref{eq4}) and (\ref{eq5}), the evaporative flux $J_{flux}$ is non-uniform across the surface and depends on the wettability of the substrate.  

\begin{figure}
    \centering
    \includegraphics[width =\linewidth]{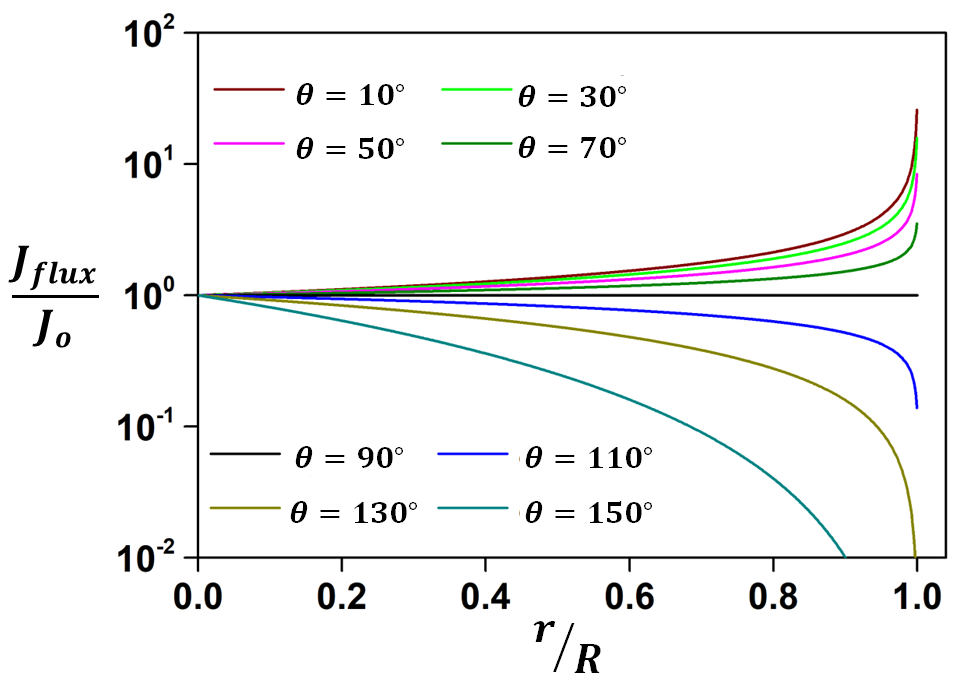}
    \caption{Normalised local diffusive evaporative flux ($J_{flux} /J_0$) over the droplet surface as a function of normalised radial distance ($r/R_d$) for various $\theta$ ($10^\circ$ to $150^\circ$).}
    \label{fig_flux}
\end{figure}

\ks{The variation in evaporative flux along the drop surface with different substrate contact angle ($\theta$) is depicted in Fig.~\ref{fig_flux}.} Depending on the value of $\theta$, the substrates are classified as - (i) hydrophilic substrate ($\theta < 90^{\circ}$), (ii) neutrally wetting substrate ($\theta = 90^{\circ}$), (iii) hydrophobic substrate ($\theta > 90^{\circ}$) and (iv) super-hydrophobic substrate ($\theta > 150^{\circ}$). From Fig.~\ref{fig_flux}, it is apparent that for a sessile drop with $\theta < 90^{\circ}$ i.e. hydrophilic substrates, $J_{flux}$ is maximum at the contact line and tend to diverge. While at the apex, $J_{flux}$ is minimum and attains a constant value. In contrast, for the hydrophobic or superhydrophobic substrates i.e. for $\theta > 90^{\circ}$, it is the reverse, $J_{flux}$ is maximum at the apex while the minimum at the edge. Interestingly, when $\theta = 90^\circ$, i.e. for a neutral wetting substrate, $J_{flux}$ is constant and is uniform over the surface of the droplet. 

\subsubsection{Free convection model} \label{sec:theory_drop_conv}
The evaporation of a droplet on a heated substrate depends on both the Stefan flow and the natural convection. \citet{kelly2018correlation} developed a relationship that accounts for both the diffusive and the concentration driven natural convective mass fluxes from an evaporating sessile droplet; albeit for unheated substrates. Thus, \citet{gurrala2019evaporation} incorporated a modified version of this relationship to incorporate the influence of a heated substrate. The evaporation mass transport rate due to the free convection derived by \citet{gurrala2019evaporation} is given by
\begin{equation}\label{eq_conv}
\left(\frac{dm}{dt}\right)_c = h_m A_s (\rho_{v,s} - \rho_{v,\infty}),
\end{equation}
where $h_m$ is the convective mass transfer coefficient, $\rho_{v,s}$ is the density of the air-vapour mixture just above the droplet free surface and $\rho_{v,\infty}$ is the density of the ambient medium. The liquid-vapour interface area, $A_s$ is given by (assuming a spherical cap profile)
\begin{equation}
A_s = {2 \pi R^2 \over 1 + \cos \theta}.
\end{equation}
Neglecting the Stefan flow, the total mass transport rate can be calculated as
\begin{equation}\label{eq_massrate}
\left(\frac{dm}{dt}\right)_{d+c} = \left(\frac{dm}{dt}\right)_d + \left(\frac{dm}{dt}\right)_c,
\end{equation}
where the first and second terms in the right hand side {of Eq. (\ref{eq_massrate})} are the diffusion mass transfer rate and the free convective mass transfer rate, respectively.

The convective mass transfer coefficient is usually expressed in terms of the convective Sherwood number, $Sh_c \equiv  {h_{m}R/ {\cal D}}$. To incorporate the convective and the diffusive mass transfers in a single expression, we can define a diffusion Sherwood number, $Sh_d \equiv {h_{d}R/{\cal D}}$. The diffusive mass transfer coefficient, $h_{d}$ can then be calculated from the following relation
\begin{eqnarray}
\left(\frac{dm}{dt}\right)_d &=& \pi R {\cal D} \mathcal{M}(c_{sat}(T_s) - c_{\infty}(T_{\infty}))f(\theta) \nonumber \\ &=& h_{d} A_s \mathcal{M}(c_{sat}(T_s) - c_{\infty}(T_{\infty})),
\end{eqnarray}
where $\mathcal{M}$ is the molecular weight of liquid vapour, $c_{sat}(T_s)$ and $c_{\infty}(T_{\infty})$ are the saturated vapour concentration at the substrate temperature and the vapour concentration in the ambient. The expression for the diffusion Sherwood number is given by 
\begin{equation}\label{eq_Sh_diff}
Sh_d = \frac{f(\theta)(1+\cos \theta)}{2}.
\end{equation}
When both the diffusion and convective fluxes are present, the effective Sherwood number is given by  
\begin{equation}\label{eq_Sh_corr}
Sh_{cor} = Sh_{d}^{*} + Sh_{c}^{*},
\end{equation}
where $Sh_{d}^{*}$ and $Sh_{c}^{*}$ are the modified diffusion and the modified convective Sherwood numbers, respectively.

The correlations for the modified Sherwood numbers are given by~\cite{kelly2018correlation}
\begin{equation}
Sh_{d}^* = Sh_d \left[1+a\left(\frac{gR_{0}^{3}}{\nu_{o}^{2}}\right)^{-b}\left(\frac{\rho_{m}-\rho_{a}}{\rho_a}Sc\right)^{c-b}Ra^b\right], \label{315}
\end{equation}
\begin{equation}
Sh_{c}^{*}=d\left(\frac{gR_{0}^{3}}{\nu_{o}^{2}}\right)^i\left(\frac{\nu_o}{\nu}\right)^{2(i-j)}\left(\frac{\rho_{m}-\rho_{a}}{\rho_a}\right)^{f-j}Sc^{e-j}Ra^j,\label{316}
\end{equation}
where $R_0$ is a nominal drop radius; $\rho_m$ and $\rho_a$ denote the density of the air-vapour mixture near the interface and of ambient air, respectively; $\nu_o$ is the ambient air viscosity at the standard condition, i.e $25^\circ$C and one atmospheric pressure; $\nu$ is the viscosity of the liquid vapour at $(T_s + T_\infty)/2$; $g$ is the acceleration due to gravity; $Sc (\equiv {\nu / {\cal D}})$ is the Schmidt number of the vapour at $(T_s + T_\infty)/2$. The Rayleigh number, $Ra$ associated with the convective mass transfer is defined as
\begin{equation}
Ra=GrSc = \left(\frac{\rho_{m}-\rho_{a}}{\rho_a}\right)\left(\frac{g{R}^3}{\nu^2}\right)\times\left(\frac{\nu}{\cal D}\right),\label{eq_Ra}
\end{equation}
where $Gr$ is the Grashof number\cite{kelly2018correlation}. Once the value of $Sh_{cor}$ is evaluated from Eq. (\ref{eq_Sh_corr}), the combined evaporation mass transfer rate due to  diffusion and convection from the droplet interface can be evaluated from the following relation
\begin{equation}
\left(\frac{dm}{dt}\right)_d + \left(\frac{dm}{dt}\right)_c = h_{d+c}A_s\mathcal{M}(c_{sat}(T_s) - c_{\infty}(T_{\infty})),\label{eq_conv+diff}
\end{equation}
where the combined diffusion and convection mass transfer coefficient, $h_{d+c}$ is given by
\begin{equation}
h_{d+c} = \frac{Sh_{cor}\mathcal{D}}{R}.\label{eq_h_d+f}
\end{equation} 
The above expression (Eq. (\ref{eq_h_d+f})) can be used for the evaporation of pure liquid droplets at room temperature.

\subsubsection{Passive transport due to free convection of air} \label{sec:theory_drop_transp}
The relation considered so far incorporates the effects of both the diffusive mass flux and the convective mass flux due to the gradients in vapour concentration. However, in the case of a sessile droplet on a heated substrate, additional terms arise due to the temperature gradient driven free convection of air. This transport mass flux, denoted by $\left(dm/dt\right)_{t}$, can be expressed in terms of the mass flow rate of air as 
\begin{equation}
\left(\frac{dm}{dt}\right)_{t}=Y_{v}^s \left(\frac{dm}{dt}\right)_{a},
\end{equation}
where $Y_{v}^s$ is the mass fraction of vapour above the free surface of the droplet. The mass convection of air over the area of the heated substrate covered by the droplet can be expressed as
\begin{equation}
\left(\frac{dm}{dt}\right)_{a} = h_{m}^{a} \pi R^2 \frac{\mathcal{M}_a}{{\cal R}_u}\left(\frac{p_{\infty}^{a}}{T_{\infty}} - \frac{p_{s}^{a}}{T_{s}}\right).
\end{equation}
Here, air has been approximated as an ideal gas; ${\cal R}_u$ is the universal gas constant; $h_{m}^{a}$ denotes the mass transfer coefficient for air; ${\cal M}_a$ is the molecular weight of air; $p_{\infty}^a$ and $p_{s}^{a}$ are the partial pressures at the ambient and substrate, respectively. The Sherwood number for air is given by $Sh_{a} = (h_{m}^{a} R)/{\cal D}_{a}$, wherein ${\cal D}_a$ is the diffusion coefficient of air. For natural convection over a horizontal flat surface, $Sh_a$ is related with the air Rayleigh number, $Ra_a$ as follows \citep{lloyd1974natural},
\begin{equation}
Sh_a=0.54Ra_a^{1/4},
\end{equation}
where 
\begin{equation}
Ra_a=Gr_a Sc_a = \left(\frac{\rho_{a} (T_s) -\rho_a (T_\infty)}{\rho_a (T_\infty)}\right)\left(\frac{g{L}^3}{\nu_a^2}\right)\times\left(\frac{\nu_a}{\cal D}_a\right).
\end{equation} 
Here all the properties are with respect to air; the characteristic length, $L = A_p/{\cal P}_p$, wherein $A_p$ and ${\cal P}_p$ are the area and perimeter of the heated substrate, respectively.

The total mass evaporation rate for liquid from a substrate at elevated temperature is thus the sum of diffusion, convection and passive transport terms given by 
\begin{eqnarray}\label{eq_ethanolhightemp}
\left(\frac{dm}{dt}\right) &=& \left(\frac{dm}{dt}\right)_d + \left(\frac{dm}{dt}\right)_c + \left(\frac{dm}{dt}\right)_t \nonumber \\ &=& h_{d+c}A_s\mathcal{M}(c_{sat}(T_s) - c_{\infty}(T_{\infty})) + Y_{v}^s \left(\frac{dm}{dt}\right)_{a}
\end{eqnarray}
and is evaluated for a pure liquid droplet at a given substrate temperature, $T_s$. The liquid density is calculated by assuming that the temperature of the droplet is equal to the substrate temperature. The calculated density is used to compute $V/V_0$ at various substrate temperatures. Here, $V_0$ and $V$ are the initial and instantaneous droplet volume, respectively. 

\begin{figure}
\centering
(a)  \\
\includegraphics[width=0.4\textwidth]{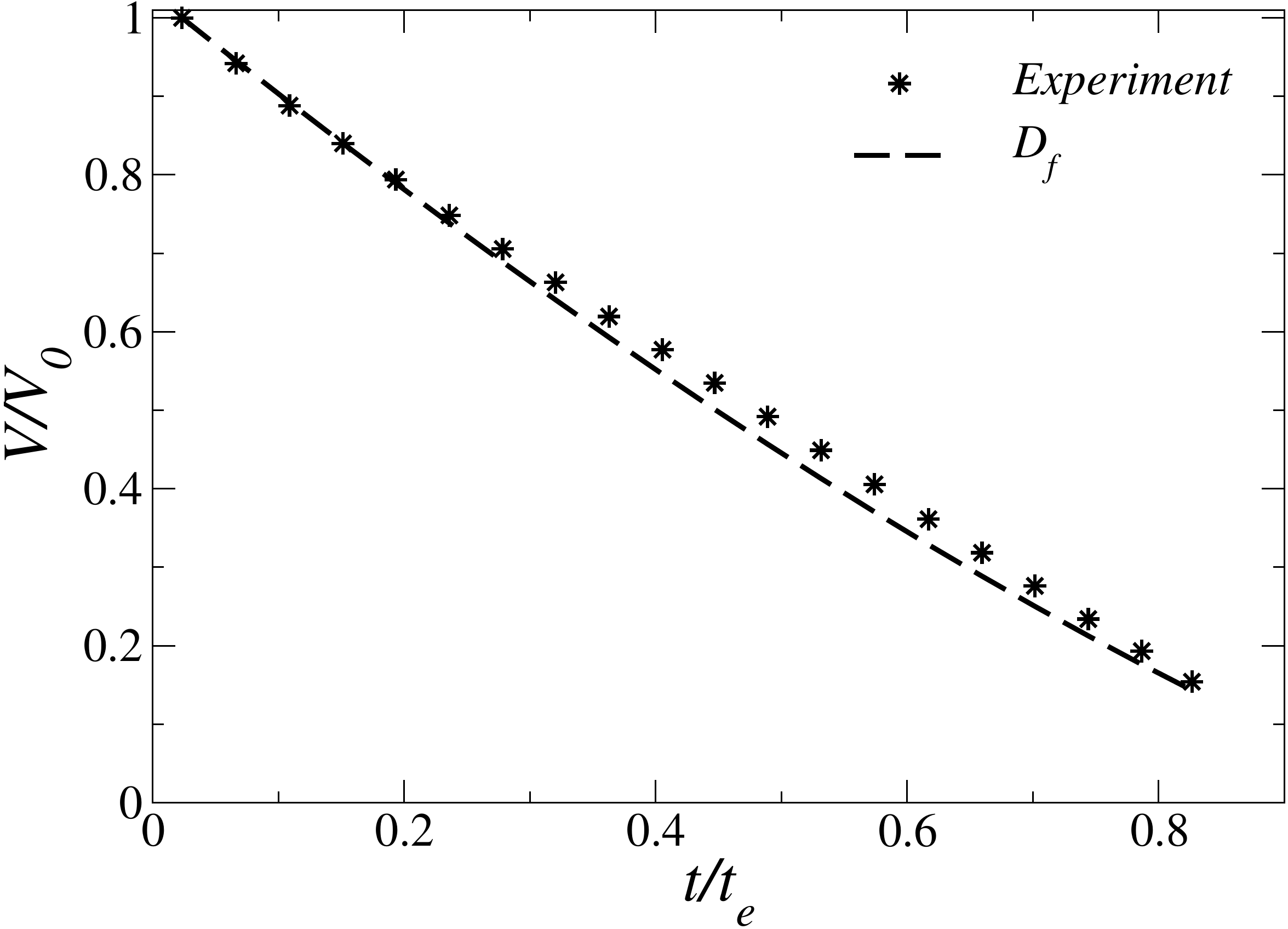}  \\
(b)  \\
\includegraphics[width=0.4\textwidth]{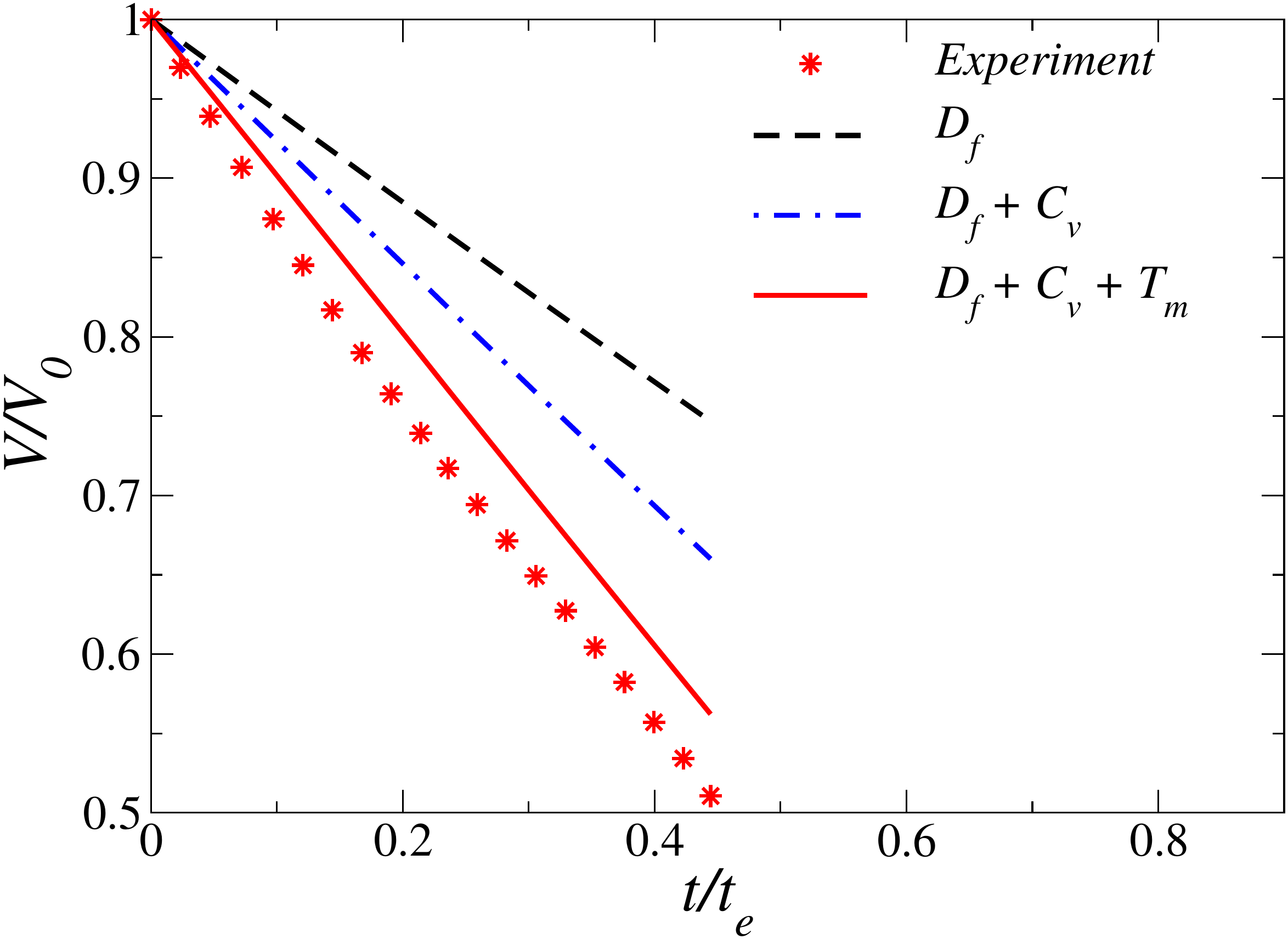} 
\caption{Comparison of the experimental and theoretically obtained $\left ({V / V_0} \right)$ versus $t/t_e$ at $T_s=60^\circ$C. Here, $t_e$ is the lifetime of the droplet. (a) Pure water droplet calculated using the diffusion ($D_f$) model alone. (b) Pure ethanol droplet calculated using diffusion ($D_f$), diffusion + convection ($D_f+C_v$) and diffusion + convection + transport ($D_f+C_v+T_m$) models.}
\label{fig:vapourpressureT25}
\end{figure}

\citet{gurrala2019evaporation} demonstrated the variations of instantaneous normalised volume $(V/V_0)$ for pure water and pure ethanol droplets, as depicted in Fig. \ref{fig:vapourpressureT25}(a) and (b), respectively. It can be observed that while the diffusion model alone shows a good agreement with the experiment for pure water droplet, it does not predict the evaporation dynamics for pure ethanol droplet. This is owing to the faster evaporation of ethanol due to its higher volatility as compared to water. \citet{gurrala2019evaporation} compared the experimental results with the models associated with diffusion (eq. \ref{case1a}), diffusion + convection (eq. \ref{eq_massrate}) and diffusion + convection + transport (eq. \ref{eq_ethanolhightemp}) in Fig. \ref{fig:vapourpressureT25}(b). It can be seen that while theoretical prediction using the diffusion model alone deviates significantly from the experimental data, including the convection effect shifts the prediction towards the experimental finding. Further, the combined diffusion + convection + transport model converges with the experimental observation very well.
 
\subsection{Evaporation of binary liquid droplets} \label{sec:theory_drop_binary}

In contrast to pure liquids, in multi-component liquid droplets, the volatility and mole-fraction of individual components vary during evaporation. Thus, the evaporation dynamics is governed by the vapour-liquid equilibrium of the binary mixtures~\cite{sandler2017chemical}. \citet{gurrala2019evaporation} experimentally investigated the evaporation dynamics of an ethanol-water mixture in a sessile droplet configuration with varying compositions and substrate temperatures. As shown in Fig. \ref{fig:vapourpressureT25}(a) and (b), a single-component (pure) droplet exhibits a monotonic linear evaporation. However, a binary droplet undergoes two distinct evaporation phases - the higher volatile ethanol evaporates rapidly compared to water, resulting in a nonlinear evaporation process. \citet{gurrala2019evaporation} also observed an early spreading stage, an intermediate pinned stage, and a late retreating stage of evaporation at a high substrate temperature. They demonstrated that the theoretical model employing the combined effect of diffusion, free convection and passive vapour transport with the freely convecting airflow predicts the evaporation dynamics of the binary liquid droplet at different temperatures.

They used the VLE plot \cite{sandler2017chemical} for vapour pressure and vapour phase mixture composition to calculate the instantaneous mass evaporation rate of the individual components (water and ethanol). The new molar composition of the liquid in the droplet for the next time step is calculated subsequently and is used in conjunction with the VLE diagram to evaluate the new vapour pressure and the bubble point composition. This iterative process is continued till the end of evaporation. The liquid solution density is evaluated at every time step, which is used to calculate the instantaneous droplet volume as shown below.
The total mass of the droplet at any instant, $m_{droplet}(t)$ is given by 
\begin{equation}
m_{droplet} (t)  = m_w(t) + m_{e} (t), \label{case1c}
\end{equation}
where $m_w(t)$ and $m_e(t)$ are the masses of water and ethanol present in the droplet of the binary mixture at any instant $t$. 
The mass fractions of water $Y_w (t)$ and ethanol $Y_e (t)$ in the droplet at any time, $t$ are given by
\begin{eqnarray}
Y_w (t) &=& {m_w (t)  \over m_w (t)  + m_e (t)} , ~~ {\rm and} \\ Y_e (t) &=& 1 - Y_w (t),
\end{eqnarray} 
respectively.  The mole fractions of water, $\chi_w (t)$ and of ethanol, $\chi_e (t)$ can also be evaluated from the following relations
\begin{eqnarray}
\chi_w (t) &=& {m_w (t)/\mathcal{M}_w  \over m_w (t)/ \mathcal{M}_w + m_e(t)/\mathcal{M}_e} , ~~ {\rm and} \\ \chi_e (t) &=& 1 - \chi_w (t),
\end{eqnarray}
where $\mathcal{M}_w$ and $\mathcal{M}_e$ are the molecular weights of water and ethanol, respectively. The ethanol-water mixture is a non-ideal solution and requires an estimation of the excess molar volume of mixing $V_{\epsilon}$ \citep{marsh1980excess}. The density of the non-ideal mixture, $\rho_{m}$ can be evaluated from the expression,
\begin{equation}
\rho_m(t) = {\chi_w(t) \mathcal{M}_w +\chi_e(t) \mathcal{M}_e \over V_{\epsilon} + \frac{\chi_w(t)\mathcal{M}_w}{\rho_w} +\frac{\chi_e(t)\mathcal{M}_e}{\rho_{e}} }. \label{case1d}
\end{equation}
where $\rho_w$ and $\rho_e$ are the densities of water and ethanol, respectively. The values of the excess volume vary with mixture composition and mixture temperature, and are either tabulated \citep{marsh1980excess} or expressed in terms of Redlich-Kister (R-K) correlations \citep{jimenez2004effect}. \ks{To calculate} $V_{\epsilon}$, the R-K polynomial expansion coefficients are used \citep{danahy2018computing}. Subsequently, the volume of the droplet of the ethanol-water mixture at any instant is given by
\begin{equation}
V(t) = {m_{droplet} (t) \over \rho_{m} (t)}.\label{case1e}
\end{equation}
The theoretical predictions were found to predict the temporal variation of the normalised instantaneous droplet volume with its initial volume obtained from experiments for both pure and binary fluids.

\begin{figure}
\centering
\includegraphics[width=0.4\textwidth]{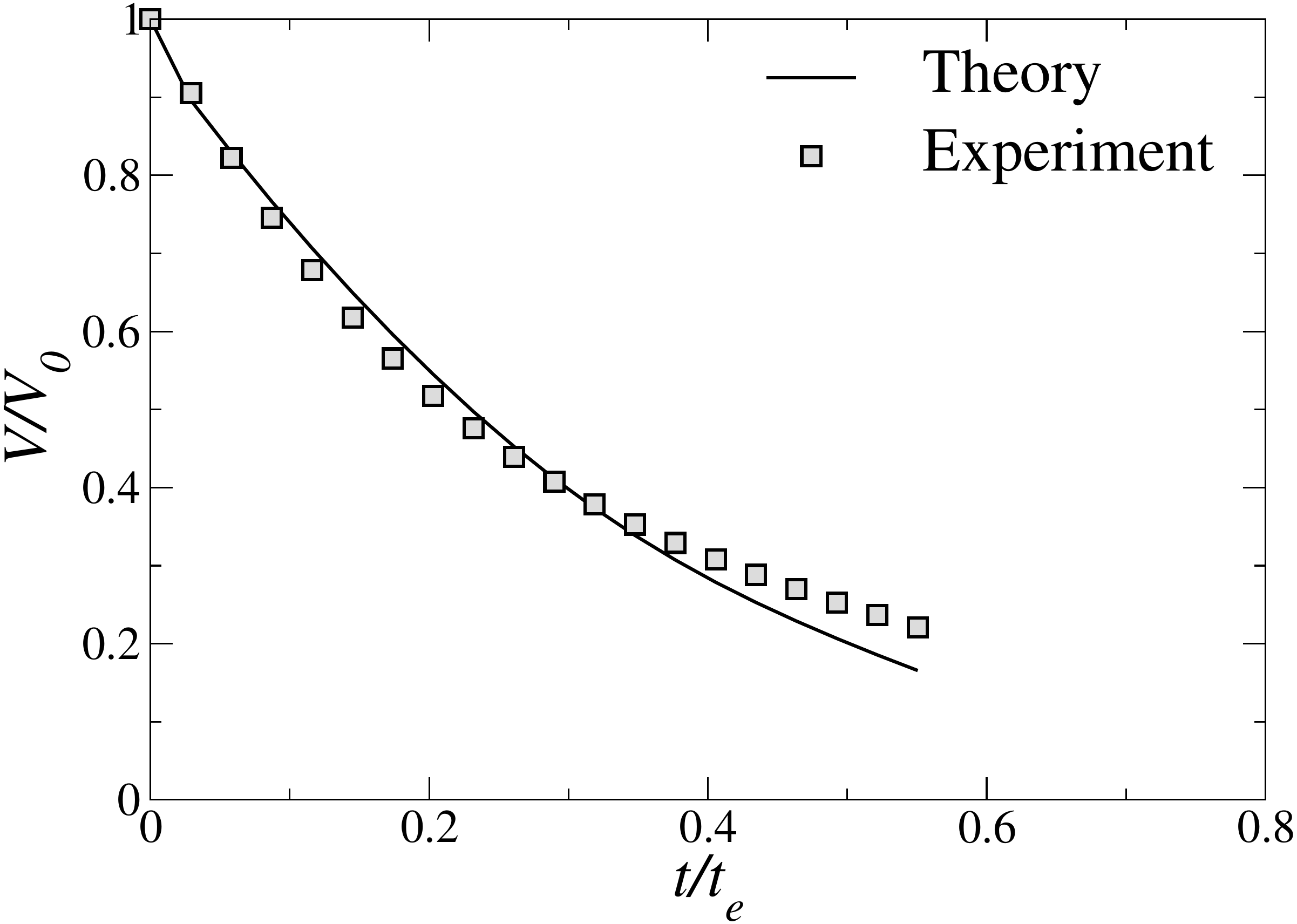}
\caption{Comparison of the experimental and theoretically obtained $\left ({V / V_0} \right)$ versus $t/t_e$ for a droplet of (ethanol 50\% + water 50\%) using diffusion + convection + transport ($D_f+C_v+T_m$) model at $T_s = 25^\circ$C \cite{gurrala2019evaporation}. \ks{Reproduced with permission from Int. J. Heat Mass Transf. 145, 118770 (2019). Copyright 2019 Elsevier.}} 
\label{fig5}
\end{figure}

Fig. \ref{fig5} compares the theoretical model considering diffusion, convection and transport with experimental data for a binary (ethanol 50\% + water 50\%) droplet. Two distinct evaporation regimes are evident due to differential evaporation rates associated with ethanol and water components. It is apparent that since ethanol is more volatile than water, the droplet initially evaporates faster, followed by slower water evaporation at a later stage.

\ks{It should be noted that the combined model discussed in the previous sections does not account for the influence of gravity. However, recent studies have revealed that even in small droplets (when the Bond number is low), gravity can still impact the evaporation process within drying drops \cite{li2019gravitational,diddens2021competing,edwards2018density}, affecting the internal flow dynamics of the droplets. Additionally, in the drying of a binary liquid drop, the density disparity between its components may induce Rayleigh convection, which competes with the Marangoni convection driven by the surface tension gradient within the droplet \cite{diddens2021competing}.}

\section{Drying colloidal drops}

\begin{figure}[htb]
\centering
\includegraphics[width=0.95\linewidth]{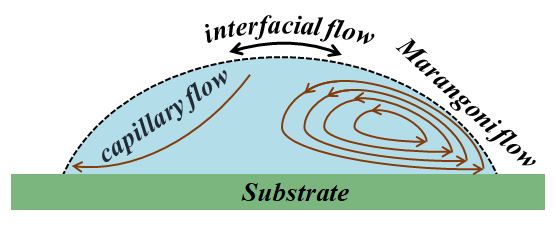}
\caption{Schematic representation of different flow fields inside the evaporating droplet. They originated via capillary flow, Marangoni flow, and the interfacial flow; arrows point to the direction of fluid flow.}
\label{Flow1}
\end{figure}

The phenomenon of drying of colloid mainly comprises (i) diffusion of vapour from the liquid-air interface to the ambient atmosphere via evaporation-induced drying, (ii) advection of the constituents and their accumulation into a particle deposit~\cite{routh2013drying, giorgiutti2018drying,hari2022counter,katre2023stability}. These phenomena have been widely studied via several standard model experiments. Here, we will discuss only the drying of sessile colloidal drops. Drying colloidal droplets \ks{leads} to the distinct pattern formation of colloids on the substrate. The formation of a colloidal deposit involves the local transportation of dispersed particles due to evaporation driven flows inside the drop. This is crucial in forming a deposit and the emergence into a distinct pattern. For a colloidal dispersion drying in a sessile configuration, the flow field inside an evaporating drop are linear~\cite{hu2005analysis,hu2006marangoni}, closed loops~\cite{hu2005analysis,hu2006marangoni}, interfacial or their combination as schematically shown in Fig. \ref{Flow1}. All of these flow fields have different origins and are known to depend on the drying geometry, the motion of the contact line, and the physicochemical condition of drying. Here, we briefly discuss a few, the most important ones that are known for the drying sessile drop - capillary flow, interfacial flow and Marangoni flow as schematically shown in Fig. \ref{Flow1}. 

\subsection{Capillary flow} 
\begin{figure*}[htb]
\centering
\includegraphics[width=0.8\linewidth]{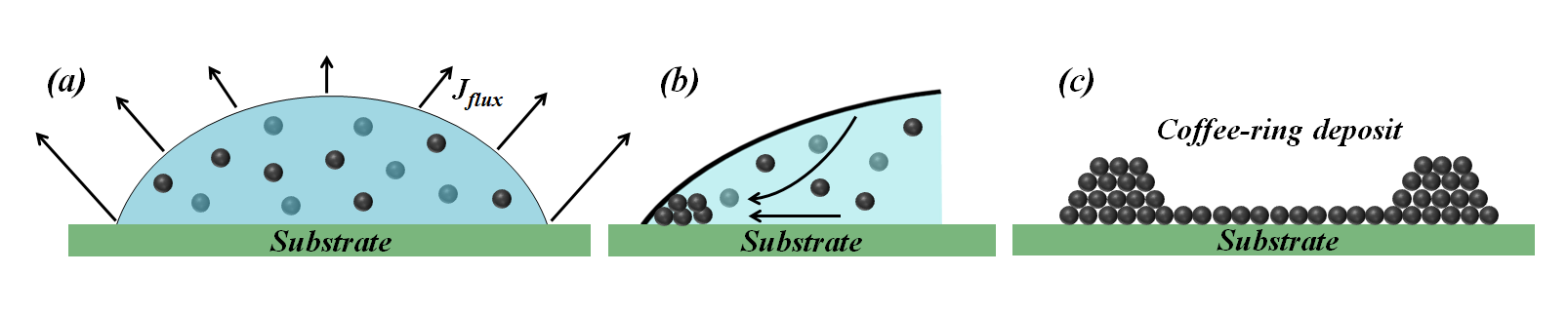}
\caption{Schematic representation of (a) the side view of an evaporating sessile drop containing dispersed particles, where arrow indicates the local evaporative flux ($J_{flux}$). (b) Transportation of particles toward the contact line represented by the arrows. (c) The side view of the final dried particulate deposit showing the accumulation of most particles at the drop edge after solvent evaporation.}
\label{fig2}
\end{figure*}

The capillary flow is omnipresent in the sessile drops drying on hydrophilic substrates. Their origin is attributed to the non-uniform evaporative flux on the droplet surface and is also known to result in a ``coffee-ring'' like deposit (Fig.~\ref{fig2}). \citet{deegan1997capillary,deegan2000contact} presented the first quantitative analysis and explained the formation of such ring-like deposits. The origin of the coffee-ring deposit was ascribed to the pinning of a contact line and the non-uniform evaporative flux ($J_{flux}$) that sets across the droplet surface. The magnitude of $J_{flux}$ varies such that their value is maximum at the contact line and minimum at the apex (depicted in Fig.~\ref{fig2} (a)). This non-uniformity in $J_{flux}$ generates an outward radial flow named \textit{capillary flow} from the bulk of a drop towards the contact line (indicated in the schematic shown in Fig. \ref{fig2} (b)). The capillary flow compensates for the higher fluid loss from the contact line and transports the dispersed particles towards the droplet edge. On complete drying, the accumulated particles manifest into a thick ring-like deposit that comprises closely packed particles as shown in Fig.~\ref{fig2} (c). The thickness, width and radius of this ring-like deposit constituting colloidal particles are known to depend on several physical parameters including initial volume fraction of the dispersion, wettability and the drying conditions~\cite{deegan2000contact}, which will be discussed in the following section. 
   
The particle migration velocity due to the outward capillary flow at any radial location ($r$) can be estimated using the following expression \cite{marin2011order},
\begin{equation}
U_c \sim \frac{D^*}{\theta(t) \sqrt[]{R(R-r)}},
\label{eq7}
\end{equation}
where $D^* = 2\sqrt{2} {\cal D} \Delta c/\pi \rho$ is the effective diffusivity of the fluid in the droplet, $\Delta c$ is the difference in concentration of liquid-vapour at the droplet surface and the ambient environment, and $\rho$ is the density of fluid, $\theta(t) $\, is instantaneous three-phase contact angle and $R$\, is the contact radius of a droplet.

As evident from Eq.~(\ref{eq7}), the capillary flow velocity is inversely proportional to $\theta(t)$ and varies with the radial distance $r$ within the droplet. This resembles the higher velocity of the fluid at the region close to the droplet contact line and is equivalent to a situation of squeezing. Note that towards the tail end of drying, i.e., when $\theta(t) \rightarrow 0$, the capillary flow velocity ($U_c$) diverges and the flow field becomes maximum which is independent of $r$. As mentioned previously, the strength of this capillary flow greatly depends on the variation in evaporative flux $(J_{flux})$ along the drop surface, which is further known to depend on the substrate wettability and evaporation rate. 

\subsection{Interfacial flow}

\begin{figure}[htb]
\centering
\includegraphics[width=0.9\linewidth]{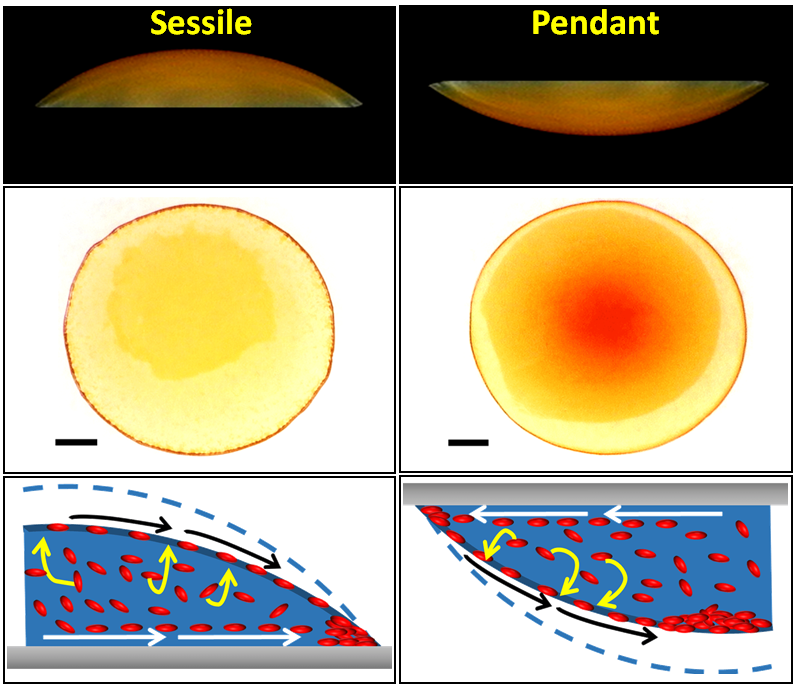}
\caption{Distinct dynamics observed in sessile and pendant drop configurations. In sessile configuration, both the capillary flow and interfacial migration are directed towards the contact line. In pendant configuration, the interfacial migration is directed towards the apex of the drop while the capillary flow is directed towards the contact line \cite{mondal2018patterns}. \ks{Reproduced with permission from Langmuir 34, 11473–11483 (2018). Copyright 2019 American Chemical Society.}} 
\label{Flow}
\end{figure}

Interfacial flow is another important advection mechanism common in drying colloidal systems. This originates from the gravity-induced deformation of a drop comprising colloidal particles~\cite{mondal2018patterns}. In the drying dispersion, particularly the sessile drop, the particles are adsorbed to the air-liquid interface during the evaporation and get trapped. These trapped particles at the interface require high energy for their detachment and prefer to remain there instead of getting detached. To reduce the interfacial energy, the drying droplet chooses an alternate route where the adsorbed particles transport along the air-liquid interface to the region of higher surface curvature~\cite{cavallaro2011curvature}. This, in turn, enhances the accumulation of particles at the edge in sessile drop drying. These flow mechanisms are prevalent for the substrates where the $\theta \geq 90^\circ$. The interfacial flow-driven accumulation mainly depends on the mean and deviatoric curvatures of the air-liquid interface~\cite{cavallaro2011curvature}. \citet{mondal2018patterns} demonstrated the drying of colloidal droplets in the sessile and pendant configurations. They showed that drying sessile droplets leads to the formation of coffee-ring like deposit, whereas, drying in pendant configuration resulted in central accumulation of particles, as shown in Fig. \ref{Flow}. In sessile configuration, the particles absorbed at the interface migrate towards the contact line due increasing surface curvature. Therefore, the particle migration along the interface is an additional transport mechanism that enhances the coffee-ring formation in the sessile configuration (Fig. \ref{Flow}). In contrast, in the pendant configuration, the capillary flow drives the particles toward the contact line, and the interfacial migration of particles is directed toward the apex of the droplet. This results in a central deposit in addition to the coffee-ring effect (Fig. \ref{Flow}).

\begin{figure*}
\centering
 \hspace{0.02cm}  (a) \hspace{4.6cm} (b) \hspace{4.6cm} (c) \\
 \includegraphics[width=0.3\textwidth]{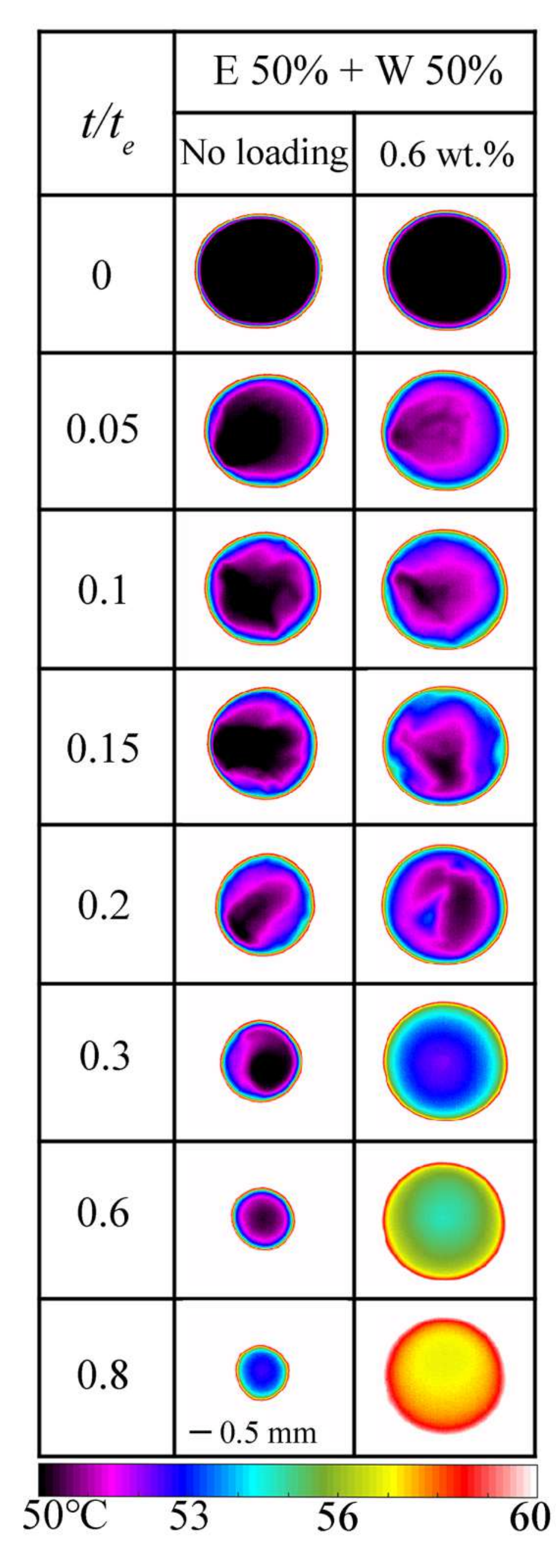} \hspace{1.0mm} \includegraphics[width=0.3\textwidth]{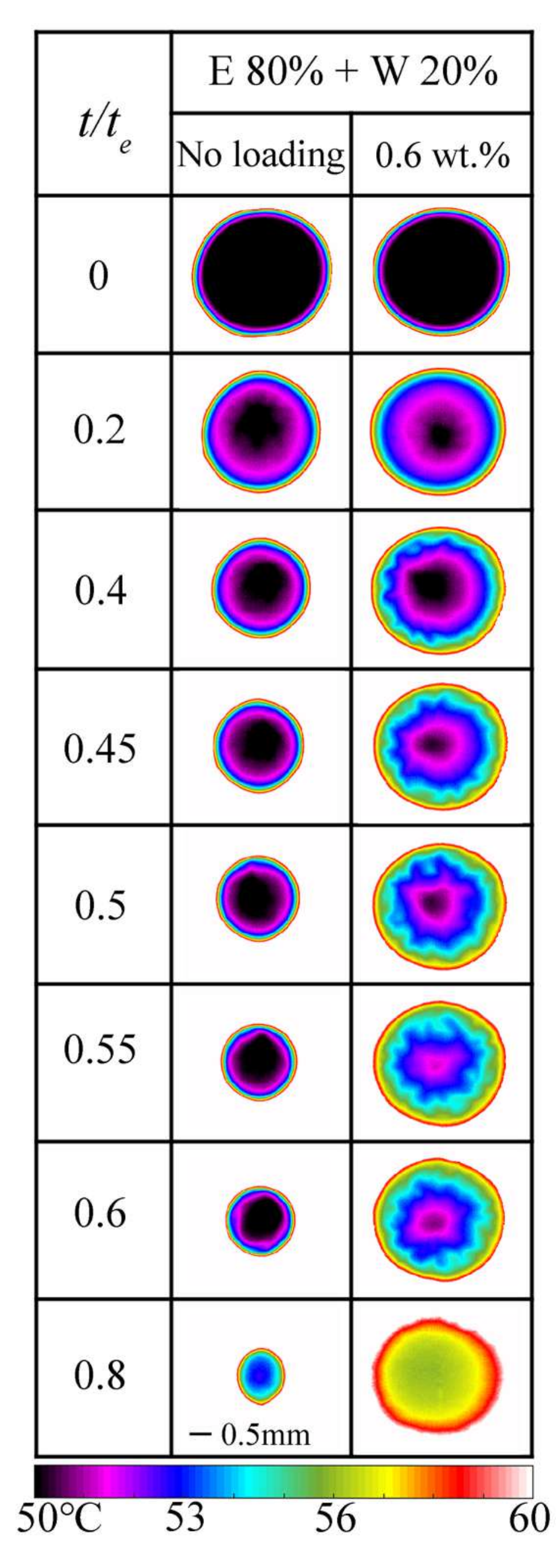} \hspace{1.0mm}
 \includegraphics[width=0.3\textwidth]{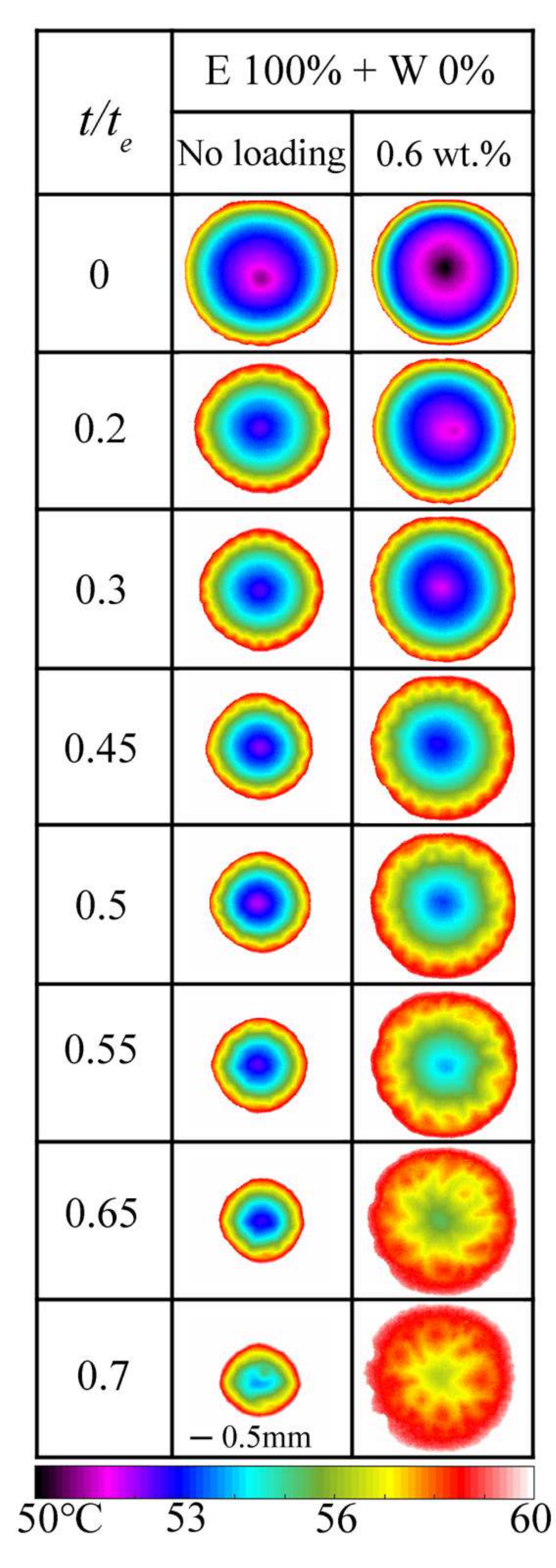}
\caption{Temporal evolutions of the temperature contours of (a) (E 50\% + W 50\%), (b) (E 80\% + W 20\%) and (c) (E 100\% + W 0\%) droplets with and without nanoparticle loading at $T_s=60^\circ$C \cite{katre2021evaporation}. The color-bars liquid-vapor temperature variations inside the droplet. \ks{Reproduced with permission from Langmuir 37, 6311–6321 (2021). Copyright 2021 American Chemical Society.}}  
\label{fig:fig9}
\end{figure*}

\subsection{Marangoni flow} 

The Marangoni flow inside a drying drop is generated due to the gradient in the surface tension along the drop surface. Typically, this flow is directed from the lower surface tension to higher surface tension region. This flow can be induced either by generating the temperature gradient along the drop surface (often referred to as thermal Marangoni effect~\cite{hu2006marangoni,weon2013fingering}) or by adding the surface-active solutes such as polymers or surfactant (referred to as the solutal Marangoni effect \cite{karpitschka2017marangoni,takhistov2002complex,girard2006evaporation}). For a sessile drop evaporating at a fixed temperature, the gradient in surface tension is developed due to differential cooling along the interface. It originates from non-uniform evaporative flux along the drop surface. \ks{The temperature at the apex of a drying sessile drop is lower than at the contact line.} This local variation in temperature depends on the relative magnitude of the thermal conductivity of a substrate and the dispersion~\cite{ristenpart2007influence}. As the surface tension is inversely proportional to the temperature, therefore there is a gradient in surface tension along the droplet surface such that the magnitude of surface tension is highest at the apex and lowest at the contact line. This induces an inward flow field followed by the advection of particles in the bulk region of the drop. The velocity of this flow field is given by~\cite{weon2013fingering}, 
\begin{equation}
    U_{Ma} = \frac{1}{32} \frac{\beta \theta(t) ^2 \Delta T}{\eta}, 
    \label{eq8}
\end{equation}
where $\beta$ denotes the gradient in the surface tension, $\theta(t)$ is the instantaneous contact angle of the drop, $\eta$ 
 corresponds to the dynamic viscosity of the fluid, and $\Delta T$ represents the difference between the maximum and minimum temperature in the drop.

As evident from Eq.~(\ref{eq8}), the velocity of fluid driven by the Marangoni flow is directly proportional to $\beta$ and $\Delta T$ while inversely proportional to $\eta$. Thus increasing $\Delta T$ or $\beta$ can enhance the Marangoni flow-driven deposition of particles. Likewise, adding surface-active agents (or surfactant) to an evaporating droplet can induce solutal Marangoni convection of the particles inside the drop~\cite{cui2012suppression}. In such cases, the outward capillary flow first advects the surfactant towards the contact line, lowering the surface tension at that region compared to the droplet apex. This sets up the gradient in surface tension with a minimum at the edge and a maximum at the contact line. Therefore, this generates an inward flow of fluid together with the constituting particles towards the droplet interior~\cite{still2012surfactant}. Fig. \ref{fig:fig9} depicts the temporal evolutions of the temperature field for binary liquid drops of compositions (E 50\% + W 50\%), (E 80\% + W 20\%) and (E 100\% + W 0\%) with and without nanoparticle loading at an elevated substrate temperature of $T_s=60^\circ$C. \citet{katre2021evaporation} showed the hydrothermal waves resulted from the combination of solutal and thermal Marangoni convection using an infra-red imaging system, as shown in Fig. \ref{fig:fig9}. It can be seen that for both the loading and no-loading conditions, increasing the ethanol concentration enhances the Marangoni flow-driven convection. This infers that the interplay between these flow fields can alter the resultant dried patterns on the substrate. For instance, in a drying of particle-laden sessile drop, the dominance of capillary flow results in a coffee-ring deposit while that of Marangoni flow results in its suppression~\cite{hu2006marangoni}. 

\section{Dried deposit Pattern} 
The dried deposit emerging into distinct patterns are controlled by various intrinsic (viz. shape, size and charge of the constituting particles) and extrinsic (viz. substrate wettability and drying condition) parameters. In this section, we will briefly describe the role of these different parameters.

\begin{figure}[htb]
\centering
\includegraphics[width=0.95\linewidth]{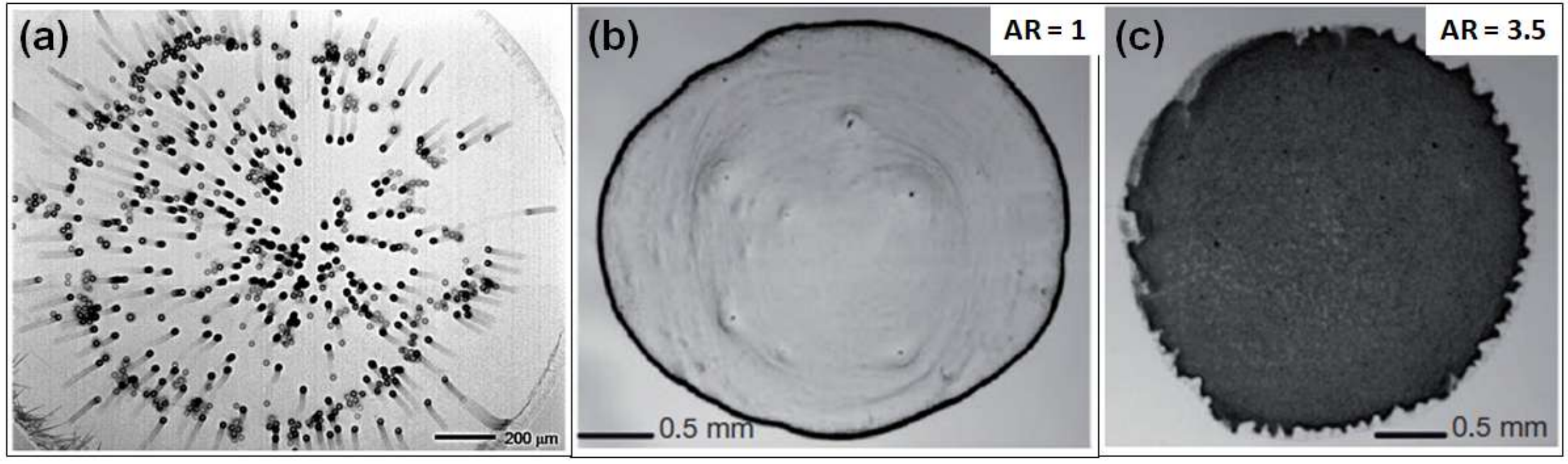}
\caption{Optical microscopy images of (a) drying droplet showing the reverse capillary motion of the dispersed particles for $10\mu$m polystyrene micro-spheres. Optical microscopy images of a dried droplet (b) with polystyrene micro-spheres show coffee-ring deposit while (c) with polystyrene ellipsoids exhibit uniform deposit \cite{zang2019evaporation}. \ks{Reproduced with permission from Physics Reports 804, 1–56 (2019). Copyright 2019 Elsevier.}} 
\label{fig4_2}
\end{figure}

\subsection{Particle size and shape} 
The size of the particle is found to play a pivotal role in controlling deposit patterns. The colloid evaporating in sessile configuration, containing smaller size particles (diameter $\sim 2\,\mu$m), predominantly forms coffee-ring-like deposits. While that comprising of bigger size particles (diameter $\sim 20\,\mu$m) forms multiple ring-like deposit (shown in Fig.~\ref{fig4_2}(a))~\cite{weon2010capillary,zang2019evaporation}. The variation in the size of the suspended particles alters the flow field inside the evaporating drop and is known to originate from the reverse capillary motion of fluid~\cite{weon2010capillary}. \ks{
In recent studies by \citet{li2018pattern} and \citet{kumar2021patterns} demonstrated that the combination of substrate orientation and particle size can greatly influence the final pattern of a dried deposit. This is due to a combination of factors including gravity settling of particles, interface shrinkage and capillary migration. Notably, for larger particles settling under the influence of gravity, a transition from a coffee-ring to a coffee-eye deposit pattern is observed as a function of orientation of the substrate.} Apart from the size of constituent particles, their shape has also been shown to alter the deposit pattern. For example, \citet{yunker2011suppression,zang2019evaporation} have shown that for a sessile drop containing a polystyrene sphere, the drying yields a coffee-ring deposit (shown in Fig.~\ref{fig4_2}(b)), while that with particle aspect ratios ($AR$) greater than 1.5 results in uniform depositions (Fig.~\ref{fig4_2}(c)). The formation of the uniform deposit is attributed to the adsorption of the ellipsoids at the water-air interface that prevents their deposition at the contact line.
\begin{figure}[htb]
    \centering
    \includegraphics[width=\linewidth]{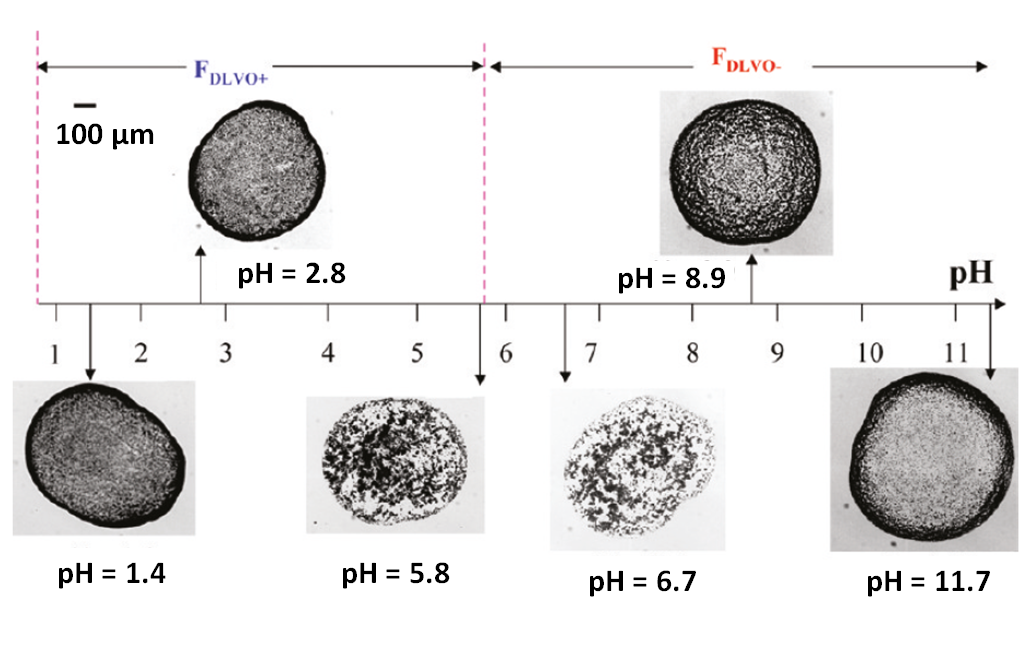}
    \caption{Images of dried droplets showing the variation in their dried patterns as a function of pH of a drying dispersion. All of these drops contain titania nanoparticles of diameter = 25\, nm possess initial concentration = 2\% ($v/v$) and were dried at temperature $25^\circ$C \cite{bhardwaj2010self}. \ks{Reproduced with permission from Langmuir 26, 7833–7842 (2010). Copyright 2010 American Chemical Society.}} 
    \label{dlvo_fig}
\end{figure}

\subsection{Particle surface charge} 

Besides the size and shape of the particles, the surface charge of the constituting particles is also found to play a crucial role in the process of consolidation of colloidal particles. In principle, colloidal particles dispersed in a continuous medium can be assumed as a dispersed dielectric sphere that experiences a repulsive electrostatic force and attractive van der Waals force (together called DLVO interaction). The surface charge $\Sigma_p$ on these particles are given by~\cite{bhardwaj2010self},
\begin{equation}
    \Sigma_p = -\frac{F\Delta n}{m_p S_p},
    \label{eq1_8}
\end{equation}
where $F$ is the Faraday's constant (96500 C/mol), $\Delta n$ is the number of ions adsorbed onto the surface of particles, $m_p$ is the mass of the particles (in kg), and $S_p$ is the specific surface area of a particle (in m$^2$/kg). The $\Sigma_p$ on particles is modulated by controlling the number of $H^+$ or $OH^-$ ions by controlling the dispersion pH. Similarly, the number of ions that are accumulated on the substrate and the net interaction force exerted on them can also be modulated by varying the pH of a solvent.

\begin{figure}[htb]
\centering
\includegraphics[width=\linewidth]{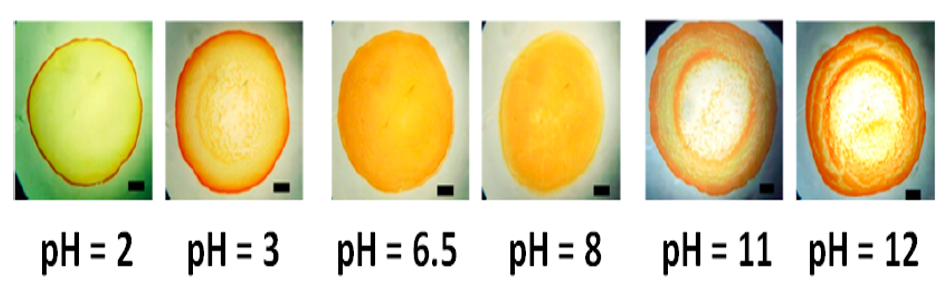}
\caption{Images of the dried droplets containing hematite ellipsoids of AR\,$=4$ (with $\rm length/breadth=200 nm/50nm$ with varying pH shows the variation in the deposit pattern. All the drops were obtained at temperature\,$=25^\circ$C and RH\,$=40$\%). The scale bar in each image corresponds to 500 $\rm \mu m$ \cite{dugyala2014control}. \ks{Reproduced with permission from Langmuir 26, 7833–7842 (2010). Copyright 2010 American Chemical Society.}} 
\label{fig6}
\end{figure}

\citet{bhardwaj2010self} have shown that for a dispersion comprising of titania particles, dried in a sessile configuration, the variation in pH results in distinct dried patterns as shown in Fig.~\ref{dlvo_fig}. At pH\,$<5$  and pH\,$> 9$, the dried deposit is coffee-ring-like, while in the range $5<$\,pH\,$<9$, deposit is uniform. The variation in deposit profile at different pH is attributed to the relative strength of interaction forces between the particles and the particle-substrate. A similar observation is reported by \citet{dugyala2014control} for the dispersion of non-spherical hematite ellipsoids as shown in Fig.~\ref{fig6}. As evident in Fig.~\ref{fig6}, a coffee-ring deposit is obtained when an aqueous drop dried at a highly acidic (pH $<5$) or highly basic pH ($>9$) conditions. \ks{However, an intermediate pH results in a uniform deposit pattern, as shown in Figure~\ref{fig6} for pH=6.5 and 8.}

\begin{figure}[htb]
\centering
\includegraphics[width=0.95\linewidth]{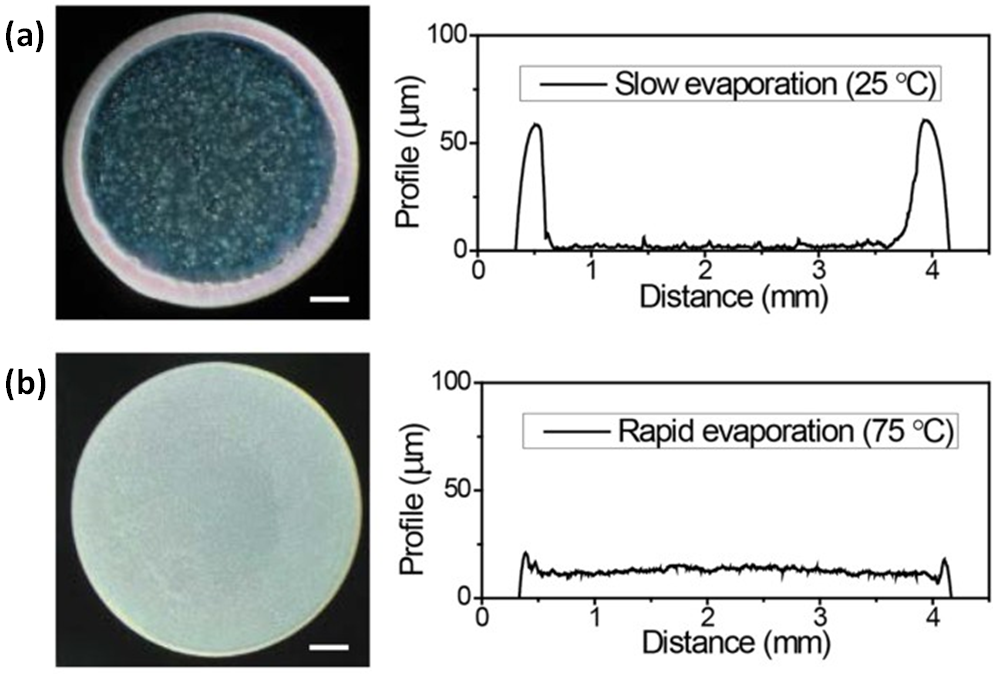}
\caption{Optical microscopy images of the dried sessile drops comprising polystyrene microspheres, obtained - (a) at room temperature $T_s$ = 25$^{\circ}$C) and (b) at an elevated temperature $T_s$ = 75$^{\circ}$C) with constant relative humidity, RH\,$\approx$\,50\%. The spatial variation in their height profiles is depicted in the right panel \cite{li2016rate}. \ks{Reproduced with permission from Sci. Rep. 6, 1–8 (2016). Copyright 2016 Springer Nature.}} 
\label{Temperature}
\end{figure}

\subsection{Drying condition} 

The environmental conditions under which the drying experiments are performed can potentially alter the nature of the deposit patterns \cite{patil2016effects,uno1998particle,zhang2013ring}. The evaporation rate and flow field significantly vary on changing the ambient condition, particularly the substrate temperature ($T_s$) and relative humidity ($\rm RH$). They directly affect the rate of diffusion of liquid vapour from the air-water interface to the ambient atmosphere and the mobility of the particles inside the dispersion. For a sessile drop of colloids comprising of polystyrene micro-spheres and drying at different temperatures, \citet{li2015coffee} have shown that the deposit can undergo the transition from a ring-like deposit to a uniform deposit (shown in Fig.~\ref{Temperature}(a)-(b)). For a sessile drop drying at a temperature $T_s \approx 25^\circ$C, the deposition is mostly governed by outward capillary flow, which results in the deposition of particles at the edge and result in ring-like deposit as shown in Fig.~\ref{Temperature}(a). While drying at a higher temperature ($T_s > 70^\circ$C), the air-water interface of a drop is found to descend at a significantly higher rate. This result in the capture of particles at the interface and hinders the capillary-flow driven deposition of particles at the edge (prevalent at $T_s \approx 25^\circ$C). Finally, a uniform deposit is obtained as shown in Fig.~\ref{Temperature}(b). \ks{The effect of pressure on pattern formation in a drying colloidal droplet was recently studied by \citet{zhang2022ultrafast}. The authors found that in a low-pressure environment, the evaporation rate increases significantly, causing particles to be captured at the interface, resulting in a uniform deposit.} Other popular schemes that are also exploited to alter the flow field and the deposited pattern are drying in the presence of (a) electric field~\cite{nobile2009self} and (b) the acoustic waves~\cite{mampallil2015acoustic}, which are not in the scope of this review.  

\begin{figure}[htb]
\centering
\includegraphics[width=\linewidth]{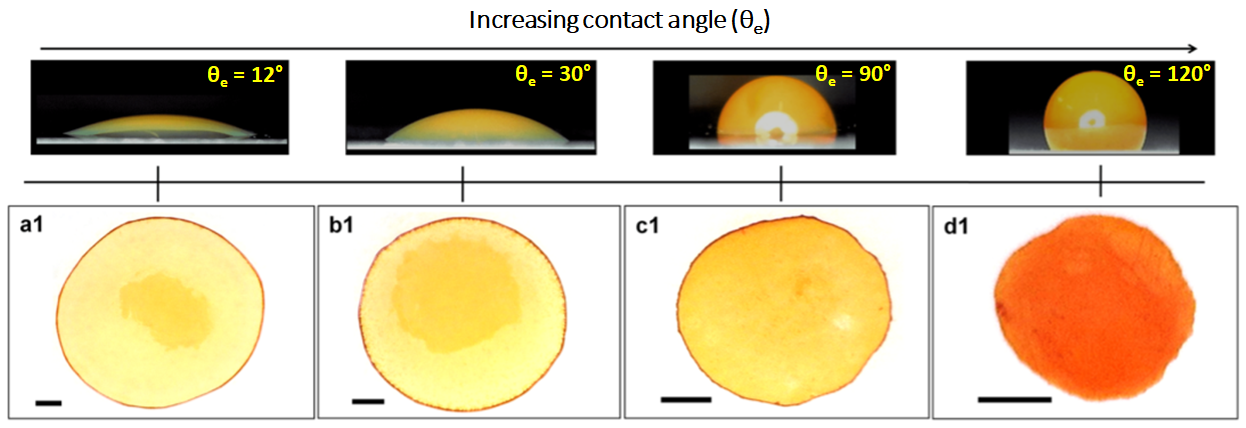}
\caption{Optical microscopy images of the dried sessile drops containing hematite ellipsoids of AR\,$=4.3$ (maintained at pH $\sim$ 2) on the substrates with various $\theta_e = [12^\circ, 120^\circ]$. All the deposits are obtained at temperature\,$=25^\circ$C and RH\,$=50$\%. The scale bar in each image corresponds to 500 $\rm \mu m$ \cite{mondal2018patterns}. \ks{Reproduced with permission from Langmuir 34, 11473–11483 (2018). Copyright 2018 American Chemical Society.}} 
\label{Wettability}
\end{figure}

\subsection{\ks{Configuration of droplets}}

\ks{The initial configuration of a drying drop can significantly alter the final dried deposit pattern. This can be manipulated by changing the substrate wettability or orientation or introducing geometrical confinement or their combination. The evaporative flux along the droplet surface strongly depends on the substrate wettability, as evident from Eq.~(\ref{eq4}) and Eq.~(\ref{eq5}).} The evaporative flux $J_{flux}$ can be seen to vary with the equilibrium contact angle $\theta$. Moreover, the uniformity and non-uniformity in $J_{flux}$ across the air-fluid interface also depend on the magnitude of $\theta$. This induces a notable change in the flow field that is generated during drying. For example, a sessile drop drying on a hydrophilic substrate, i.e. with $\theta < 30^\circ$, the dominant mode of transport is outward capillary flow. While, when $\theta > 30^\circ$, the flow field inside the drop are reported to be the closed loops (depicted in Fig.~\ref{Flow1}). They are generated via thermal Marangoni flow \cite{hu2005analysis,hu2006marangoni}. Conventionally, the drying of a colloidal dispersion is mostly studied on hydrophilic substrates with $\theta$ $<$ 90$^{\circ}$. The nature of the deposit is often ring-like, as shown in Fig.~\ref{Wettability}(a1)-(c1). The deposition of constituting particles at the contact line via evaporation-induced flow decreases with the increase in $\theta$. For a colloid drying on a hydrophobic substrate, it often yields a uniform deposit as shown in Fig.~\ref{Wettability}(d1). The possibility of generating the structured surfaces has made it possible to investigate the evaporation-induced drying studies on the hydrophobic and super-hydrophobic surfaces with $\theta$ $>$ 90$^{\circ}$ \ks{\cite{mchale2005analysis,gelderblom2011water,li2018drop}}. They often form multiple rings of particulate deposits. Further, this investigation has been extended as the function of surface roughness. They are found to significantly affect the formation of the deposit patterns and influence the spatial distribution of particles in the dried deposit~\cite{lohani2018nanoscale}.
\begin{figure}[htb]
\centering
\includegraphics[width=0.95\linewidth]{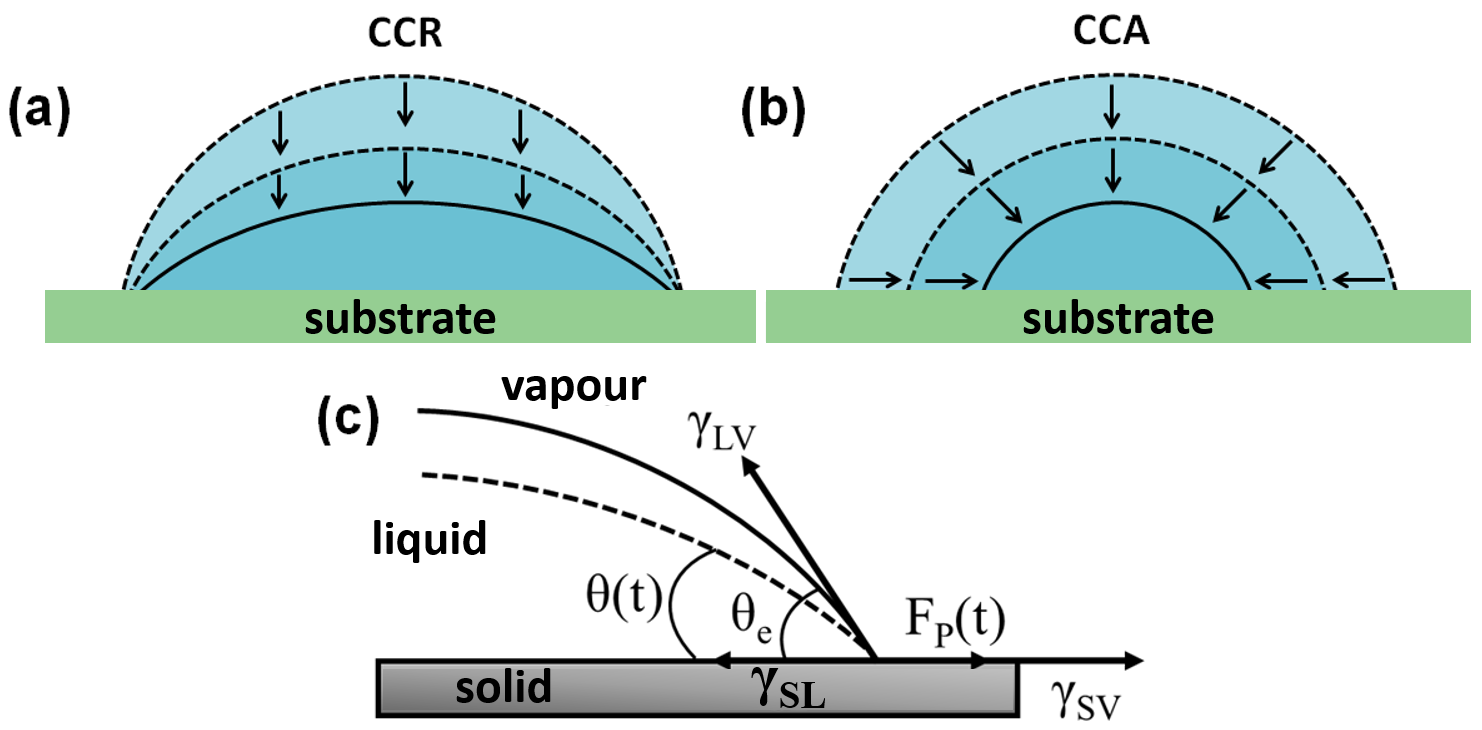}
\caption{Schematic shows the contact line dynamics of an evaporating droplet. (a) Constant contact radius (CCR) mode where the diameter of the drop remains constant during drying and (b) constant contact angle (CCA) mode where the contact angle remains the same and diameter of the drop decreases during drying. The arrows indicate the direction of receding of a droplet surface. (c) Schematic showing the force balance along the pinned three-phase contact line.}
\label{fig3}
\end{figure}
\begin{figure}[ht]
\centering
\includegraphics[width=0.95\linewidth]{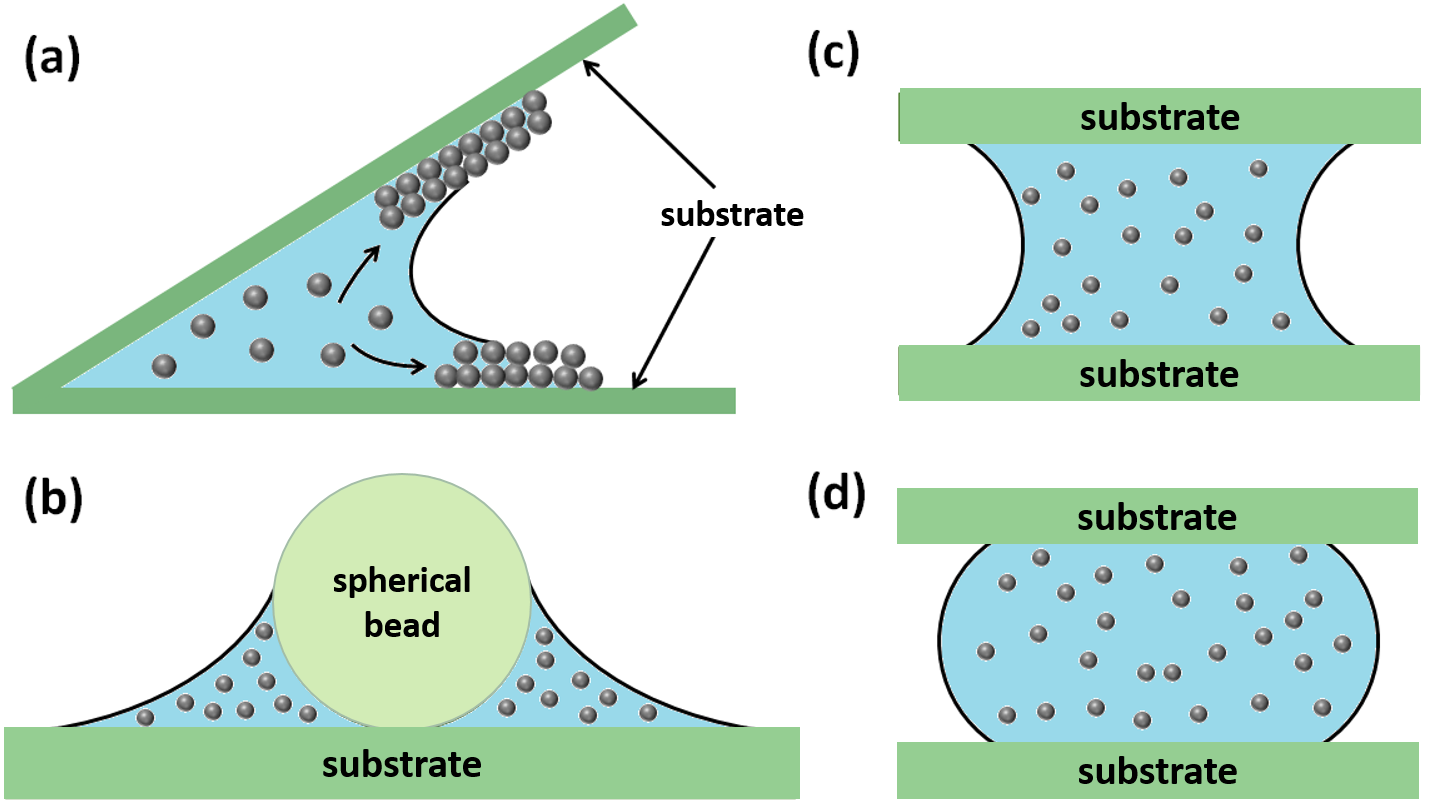}
\caption{Schematic of different drying configurations - (a) wedge geometry, (b) sphere-on-plate, (c) confined between two hydrophilic plates and (d) hydrophobic substrates.}
\label{fig7}
\end{figure}


\ks{In addition to the substrate wettability, drying colloidal dispersion under geometrical confinement is expected to gain control over the evaporation process by manipulating the contact line dynamics.} Therefore, the investigation of contact line dynamics during the process of drying is very essential to develop the methodology for controlling the evaporative-induced pattern formation~\cite{picknett1977evaporation, shanahan2001condensation}. In sessile drop drying experiments, evaporation occurs mainly in two different modes: (i) constant contact radius (CCR) mode and (ii) constant contact angle (CCA) mode. In CCR mode of evaporation, the drop evaporates such that the droplet is pinned and the contact area is constant during drying, as depicted in Fig.~\ref{fig3}(a). While in the CCA mode of evaporation, the contact angle is constant, and the contact radius decreases during the drying as shown in Fig.~\ref{fig3}(b). The evaporation process can be more complex when the two evaporation modes occur one after the other or occur simultaneously~\cite{zhang2015mixed}. The evaporation with a combination of two modes (CCR + CCA) is typically referred to as a mixed mode of drying and it results in the ``stick-slip'' motion of the contact line. Typically, for a drop evaporating on a hydrophilic substrate with $\theta$ < $90^{\circ}$, the contact line is initially pinned or ``stick'' to the substrate at an initial drying stage ($\theta(t) > \theta_R)$. Here, $\theta (t)$ is the instantaneous contact angle and $\theta_R$ is the receding contact angle. The contact line of a drop de-pin or ``slip'' when $\theta(t) \leq \theta_R$. For a pinned contact line, the force balance at the contact line yields (schematically shown in Fig.~\ref{fig3}(c)),
\begin{align}
F_P(t)= \gamma_{LV} cos \theta(t) + \gamma_{SL} - \gamma_{SV},
\label{pinning force}
\end{align}
where $F_P(t)$ is an instantaneous pinning force per unit length, $\gamma_{LV}$, $\gamma_{SL}$, and $\gamma_{SV}$ are interfacial tensions of the liquid-vapour, solid-liquid and solid-vapour interfaces, respectively. The droplet is known to de-pin when the pinning force attains its maximum possible value and $\theta (t) \approx \theta_R$. The ``stick-slip'' motion of the contact line during the drying can either be discrete or continuous. In a discrete stick-slip motion, the pinning force along the contact line is assumed to be the same everywhere and the motion of the contact line is symmetric. This directly implies that during drying the entire contact line de-pins and subsequently gets pinned to a new location simultaneously. While during the continuous stick-slip motion of a drop, the entire contact line does not de-pin symmetrically.
\begin{figure}[htb]
\centering
\includegraphics[width=0.95\linewidth]{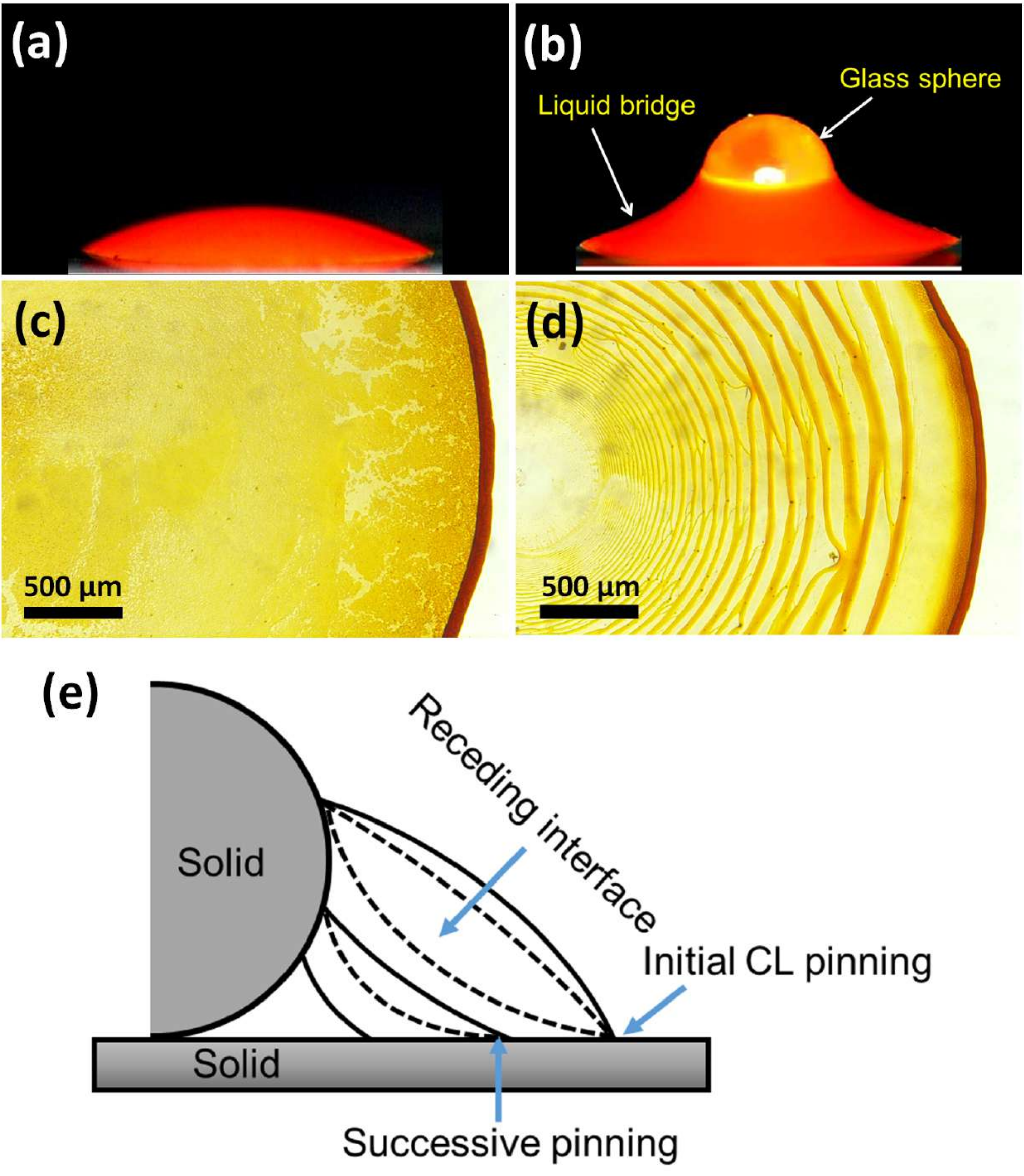}
\caption{The dispersion of hematite ellipsoids (AR\,$=5.18$) maintained at pH $\sim$ 2, drying in (a) sessile and (b) sphere-on-plate configuration. Optical microscopy images of the dried deposits obtained for suspension dried in (c) sessile and (d) sphere-on-plate configuration. (e) Schematic depicts the motion of a contact line for the dispersion drying in the sphere-on-plate configuration \cite{mondal2019influence}. \ks{Reproduced with permission from Phys. Chem. Chem. Phys. 21, 20045–20054 (2019). Copyright 2019 Royal Society of Chemistry.}}  
\label{fig8}
\end{figure}

The interplay of contact line dynamics together with the geometrical confinement has been exploited to generate various deposit patterns. Some examples of geometrically confined drying configurations are (i) wedge geometry, (ii) sphere-on-plate and (iii) confined between two parallel plates, as schematically shown in Fig.~\ref{fig7}(a)-(d). The drying of colloids inside the wedge-like geometry has been exploited to create a mono-layer assembly of colloidal particles via slow evaporation of the solvent. Similarly, the colloids dried in sphere-on-plate configuration or confined between two parallel plates are used to generate the directed self-assembly of colloidal particles resulting in novel deposit patterns. The drying of colloidal dispersion, particularly in sphere-on-plate configuration results in the formation of concentric multi-ring-like deposits, as shown in Fig.~\ref{fig8}. Such concentric multi-ring-like deposits are formed due to the discrete stick-slip motion of the contact line.   
\begin{figure}[htb]
\centering
\includegraphics[width=0.95\linewidth]{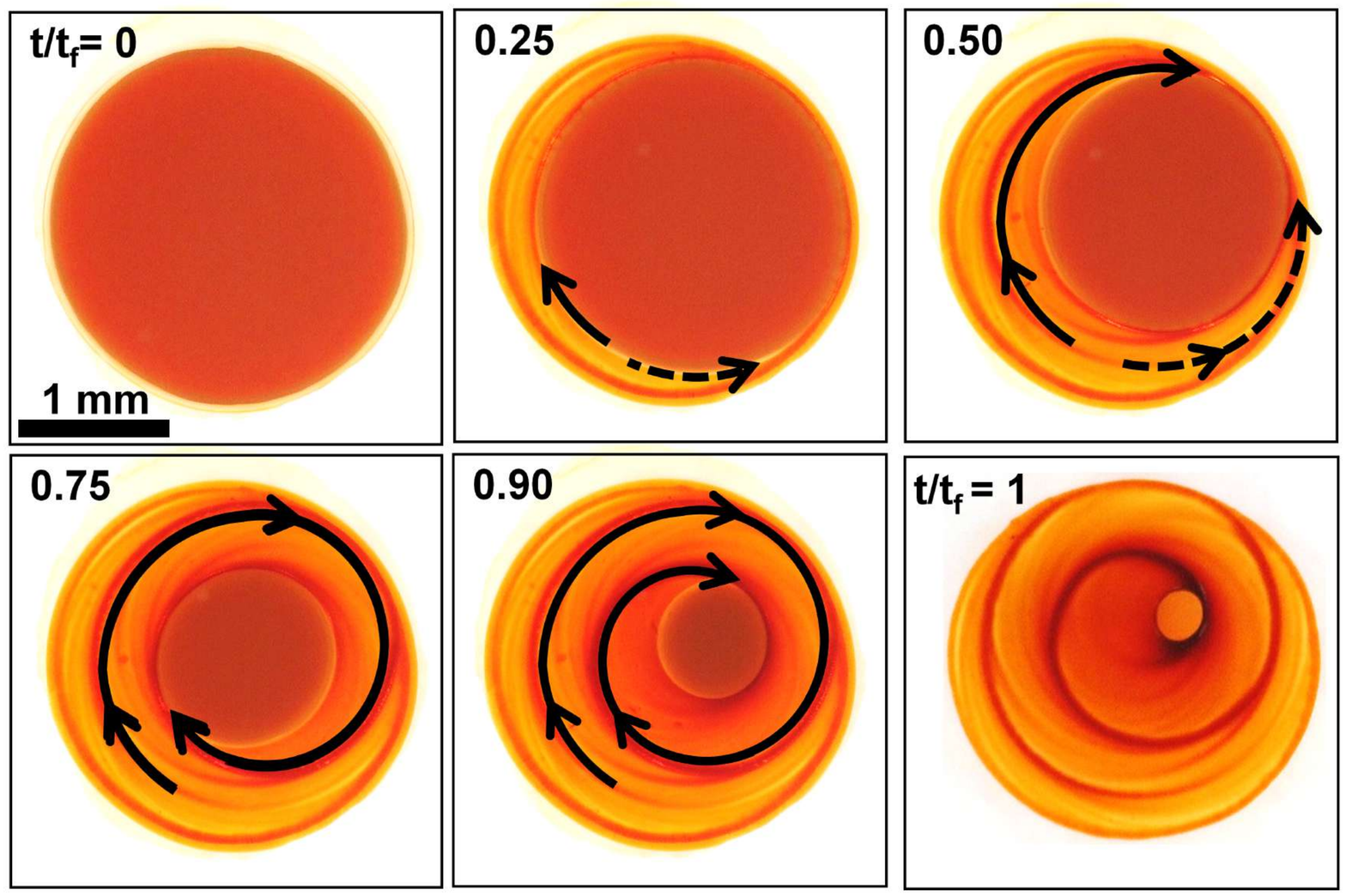}
\caption{Time-lapse optical microscopy images of the drying dispersion comprising of hematite ellipsoids that are confined between two hydrophilic glass plates. The receding motion of a contact line and its depinning are indicated by an arrow. The de-pinning of the contact line, their propagation in clockwise and anticlockwise directions, and the resultant spiral deposit are all depicted~\cite{mondal2020patterning}. The instantaneous and total drying time is given as $t$ and $t_f$ respectively. \ks{Reproduced with permission from Soft Matter 16, 3753–3761 (2020). Copyright 2020 Royal Society of Chemistry.}} 
\label{fig9}
\end{figure}
In contrast, when the pinning force varies locally along the contact line, an asymmetric motion of the contact line was observed. In those cases, the contact line remains pinned in some regions and simultaneously de-pins in other regions. This results in continuous stick-slip motion of the contact line. \citet{mondal2020patterning} reported the drying of colloids between two parallel plates. The reported deposit patterns are found to be spiral ring-like, as shown in Fig.~\ref{fig9}, which they inferred to the continuous stick-slip motion of the contact line. It is worth mentioning that the formation of such spiral deposit patterns is not affected by the direction of receding of the contact line. In other words, whether the de-pinning of the contact line is clockwise or anticlockwise or a combination of both, the resulting patterns are found to be spiral.


\section{Desiccation Cracks in dried deposit}
\begin{figure}[htb]
    \centering
    \includegraphics[width=\linewidth]{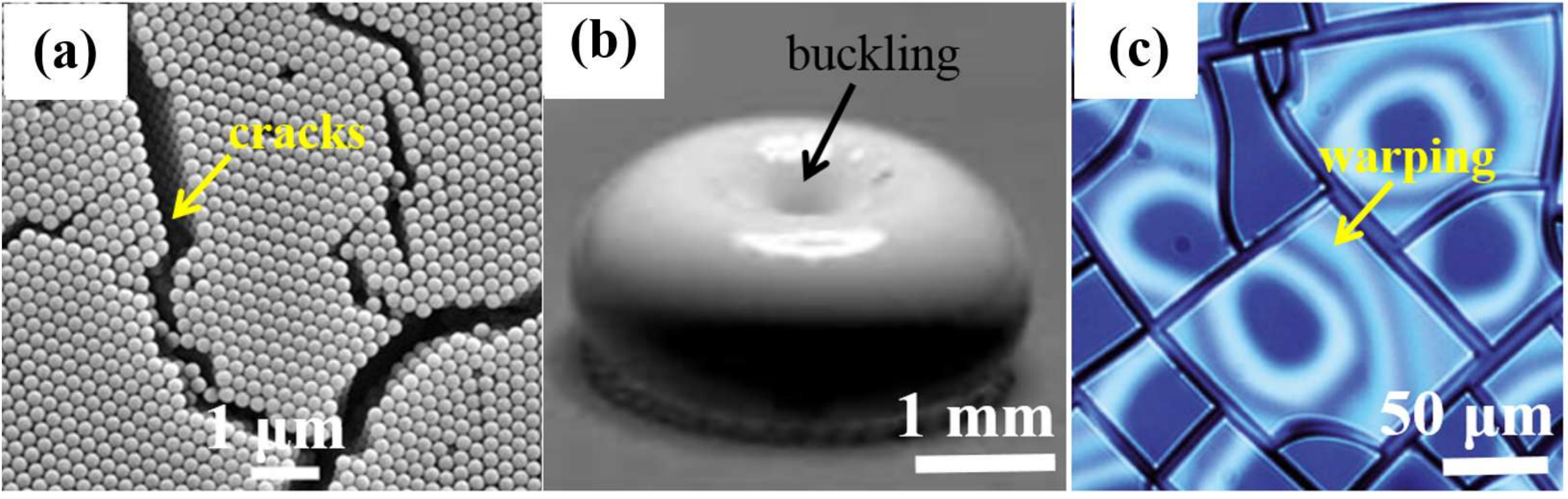}
    \caption{(a) Desiccation cracks in the deposits of poly-methyl methacrylate (PMMA) colloids \cite{hatton2010assembly}. \ks{Reproduced with permission from Proc. Natl. Acad. Sci. 107, 10354–10359 (2010). Copyright 2010 Natl. Acad. Sci. USA}. 
    (b) Colloid deposit comprising of 100\, nm latex particles shows buckling at the center \cite{pauchard2004invagination}. \ks{Reproduced with permission from Europhys. Lett. 66, 667 (2004). Copyright 2004 Institute of Physics.} 
    (c) Silica (Ludox$^\copyright$ dispersion) particulate film displays the warping as pointed by an arrow \cite{giorgiutti2015dynamic}. \ks{Reproduced with permission from Colloids Surf. A: Physicochem. Eng. Asp. 466, 203–209 (2015). Copyright 2015 Elsevier.}} 
    \label{fig_crack_1}
\end{figure}

As mentioned earlier, the drying of colloidal dispersion consolidates the particles to form a solid deposit with a close-packed structure.~\cite{routh2013drying,goehring2015desiccation} For example, when we dry a  colloidal dispersion in a sessile drop configuration, \textit{coffee-ring} like deposit is obtained. Moreover, the dried residue often accompanies cracks,  popularly known as `\emph{desiccation cracks}'. We show the images (Fig.~\ref{fig_crack_1}(a) and (c)) of desiccation cracks in dried colloidal deposits of poly-methyl methacrylate (PMMA) and silica particles, respectively. Apart from desiccation cracks, several other types of drying-induced defects, such as buckling, debonding and warping (shown in Fig.~\ref{fig_crack_1}(a-c)) are also frequently observed in the dried deposit of colloidal dispersion. It is worth mentioning that the mechanism related to nucleation and propagation of drying-induced defects is widely investigated using model experiments comprising laboratory-based drying experiments in isotropic or directional drying conditions similar to that discussed in the introduction. Here, we will only discuss the desiccation cracks. In the subsequent sections, we will briefly discuss the rudimentary concepts necessary to understand desiccation cracks. Then, we will review the literature on crack patterns and discuss the origin of different crack morphology.
\begin{figure*}[ht]
    \centering
    \includegraphics[width=\linewidth]{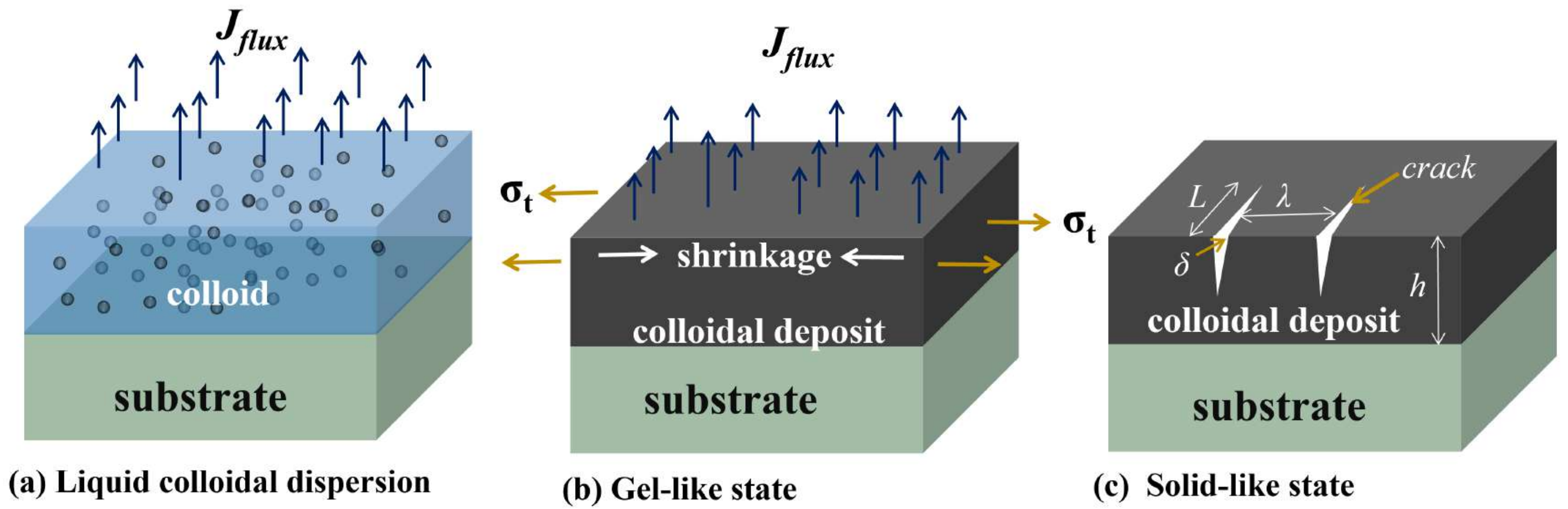}
    \caption{The various stages of the drying process of a colloidal film  resulting in the particulate deposit is schematically shown. (a) The initial stage of drying  where the colloid dispersion in the liquid state. The solvent molecules are desorbing from the surface, and the evaporative flux `$J_{flux}$' is marked. (b) Colloids are in a gel state due to the loss of fluid, causing the shrinkage of the deposit at the near-surface region. The tensile stress $\sigma_t$  generated during drying is also indicated. (c) Solid particle deposit with cracks. The crack length $L$, crack width $\delta$, crack spacing $\lambda$ and thickness $h$ are marked.}
    \label{fig_crack_2}
\end{figure*}

\subsection{Theoretical background}

The nucleation of cracks in the solid colloidal deposit occurs in several stages.~\cite{routh2013drying,goehring2015desiccation} First, the drying-induced consolidation of colloidal dispersion appears on a solid surface. Then the particulate deposits transform from a liquid state to a gel-like semisolid and then finally into solid deposits (schematically shown in Fig.~\ref{fig_crack_2}(a-c)). The loss of fluid in the gel-like state tends to shrink the top layer of the gel, while the part adhering to the substrate resists the shrinkage due to the no-slip condition. These two competing processes, namely, shrinkage of particulate deposits and resistance to the shrinkage, generate strain energy in the deposits. When the accumulated strain energy exceeds its optimum value, the crack nucleates. Please note that the generation of strain energy in the deposit also implies the modulation of associated tensile stress and strain. Citing the analogy between the cracks formed in brittle solids and the desiccation cracks, one can conceptually build up a theoretical framework for desiccation cracks. However, the origin of cracks forming in brittle solids and colloidal deposits differs. The crack nucleates from the preexisting flaw in a solid material upon applying tensile stress. In contrast, cracks in the colloidal deposits nucleate from the void between the particles due to the release of intrinsically developed stress. Furthermore, the nucleation of a crack in the colloidal deposits leads to a physical separation of particles previously bonded together during drying (Fig.~\ref{fig_crack_1}(a)). 

\subsubsection{Stress and strain} 

The stress and strain developed in a drying colloidal deposit are measurable variables that estimate the associated strain energy. They develop intrinsically due to the drying-induced shrinkage, and their magnitudes continuously modulate as drying continues. We can obtain the analytical expression for the stress and strain generated in the deposits using a continuum mechanics framework known for the \textit{poroelastic materials}. \citet{biot1941general} derived the formalism for such materials in 1941, popularly known as \textit{theory of poroelasticity}. This theory is a conjunction of fluid mechanics and structural mechanics. Assuming the drying gel-like colloidal deposits to be the isotropic linear poroelastic material, the total stress exerted on the deposits  is written as,~\cite{biot1941general,SCHERER1989171,merxhani2016introduction}
\begin{equation}
\sigma_{ij} \, = \,\tilde{\sigma}_{ij} - \alpha\,\delta_{i,j}\,\Pi,
\label{eq2_1}
\end{equation}
where $i,j$ are indices, $\sigma_{ij}$ and $\tilde{\sigma}_{ij}$ are the total stress and the hydrostatic stress exerted on the particle deposits, $\Pi$ is the pressure exerted on the void between the particles, $\alpha$ is the Biot's coefficient. The symbol $\delta_{i,j}$ is a Kronecker delta with $\delta_{i,j} = 0$ for $i\neq j$ and $\delta_{i,j} = 1$ for $i = j$.  In Eq.~\ref{eq2_1}, the first term represents the stress on the particle network, while the second term resembles the contribution of pressure exerted on the void between the particles.

The strain in the deposit can be related to the stress by Biot's constitutive relation. The constitutive relation in an index notation is ~\cite{SCHERER1989171}
\begin{equation}
    \epsilon_{ij} = \frac{1}{E_f}\Big[ (1 + \nu_f) \sigma_{ij}  - \nu_f \sigma_{ll}] - \frac{\alpha\,\Pi}{3K}\delta_{ij} 
\label{eq2_2}
\end{equation}
where, $i,j,l$ are indices, $\epsilon_{ij}$ is strain tensor, $\nu_f$ and $E_f$ are the Poison's ratio and Young's modulus of a dried colloidal deposit, $\Pi$ is the pressure exerted on the void between the particles, $K \big(= {E_f}/{3(1 - 2\,\nu_f)}$\big) is bulk modulus and $\delta_{ij}$ is Kronecker's delta function.
       
The value of $\alpha$ in Eq. (\ref{eq2_1}) and Eq. (\ref{eq2_2}) are typically considered to be $one$ for most practical purposes \cite{terzaghi1943theoretical}.

\subsubsection{Strain energy}

The strain energy per unit volume, ${U_s}_0$ accumulated in the colloidal deposit in terms of stress and strain, is given by \cite{goehring2015desiccation}
\begin{equation}
{U_s}_0 \sim \sigma_{ij} \epsilon_{ij},   
\label{eq2_3}
\end{equation}
where, $\sigma_{ij}$ and $\epsilon_{ij}$ are the stress and strain in the index notation. Further, the strain energy per unit volume in terms of free energy per unit volume ($U_{F}$) associated with the deposit is \cite{anderson2017fracture} 
\begin{equation}
    U_{F} = {U_s}_0 + U_{ex}, 
\label{eq2_4}
\end{equation}
where ${U_s}_0$ represents the magnitude of strain energy per unit volume stored in the deposit and $U_{ex}$ resembles the energy per unit volume required to create new surfaces via crack. The quantity $U_{ex}$ is related to surface-free energy by, 
\begin{equation}
    U_{ex} = 2\Gamma L,
\end{equation}
where $\Gamma$ is the associated surface free energy and $L$ is the length of the surface to be created after the crack. Since the nucleation of the crack in the colloidal deposit resembles free energy minimisation. Mathematically, this implies,
\begin{eqnarray}
\frac{dU_{F}}{dL}\Big|_c = 0 \implies \frac{d{U_s}_0}{dL}\Big|_c + \frac{dU_{ex}}{dL}\Big|_c = 0,\\
\label{eq2_5}
or, ~ -\,\frac{d{U_s}_0}{dL}\Big|_c = \frac{dU_{ex}}{dL}\Big|_c = 2\Gamma,
\label{eq2_6}
\end{eqnarray}
where the subscript `$c$' resembles the critical condition. The Eq.~(\ref{eq2_6}) implies that the crack propagates when the stored strain energy per unit volume per unit length equals the energy required for creating two new surfaces. Therefore, the extremum condition for $U_F$ gives the optimum or critical strain energy per unit volume necessary for creating new surfaces via a crack, famously called Griffith's criterion for fracture ~\cite{anderson2017fracture}. The left hand side of Eq.~(\ref{eq2_6}) is often called energy release rate or fracture energy and represented by $G_c$ \cite{anderson2017fracture}. Similarly, the quantity $d{U_s}_0/dL \equiv {d(\sigma_{ij} \epsilon_{ij})/dL}$, is known as energy release rate and symbolically represented by $-G$, where the negative sign with $G$ indicates the net decrease in strain energy via crack. Note that the quantity $G_c$ has a constant value for a given material. For a drying porous system\cite{goehring2015desiccation}, the value of $G_c$ reportedly \ks{varies from 0.1-1\,$\rm J/m^2$}. For a thin elastic plate of thickness $h$ and the Young's modulus $E_f$, under tensile stress ($\sigma_t$), the energy release rate $G$ associated with single crack is expressed as \cite{goehring2015desiccation}
\begin{equation}
    G = \frac{\pi L \sigma_t ^2}{E_f},
    \label{eq2_10}
\end{equation}
where, $L$ is the crack length as indicated in Fig.~\ref{fig_crack_2} (c). Note that using equality for energy release rates expressed in Eq.~\ref{eq2_10}, we can also obtain critical stress $\sigma_c$, where the critical stress is the minimum tensile stress that a deposit should acquire for the failure of the deposits \cite{tirumkudulu2005cracking}.

\subsection{Crack pattern}
The desiccation cracks in the colloidal deposit often self-organize and emerge into a distinct pattern \cite{groisman1994experimental}. On numerous occasions, cracks are regular,~\cite{allain1995regular, kitsunezaki1999fracture, komatsu1997pattern, inasawa2012self, Marthelot2014} or self-similar~\cite{Bohn2005}.
\begin{figure}[ht]
    \centering
    \includegraphics[width=\linewidth]{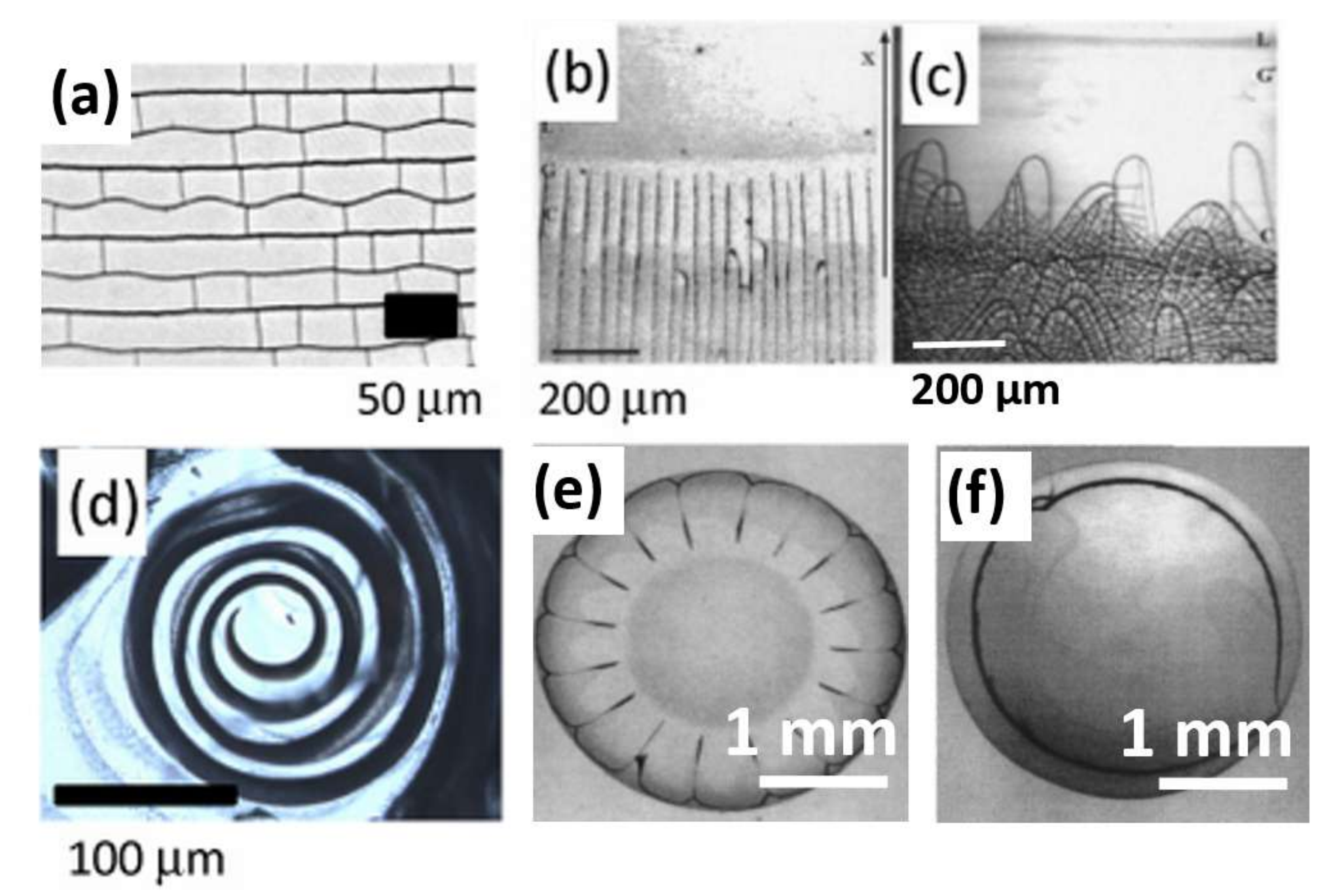}
    \caption{Various morphologies of cracks in the colloidal deposit - (a) Wavy cracks \cite{goehring2011wavy}. \ks{Reproduced with permission from Soft Matter 7, 7984–7987 (2011). Copyright 2011 Royal Society of Chemistry.} 
    (b) Linear cracks \cite{pauchard2003morphologies} and 
    (c) Arch-like crack \cite{pauchard2003morphologies}. \ks{Reproduced with permission from Phys. Rev. E. 67, 027103 (2003). Copyright 2003 American Physical Society.} 
    (d) Spiral crack \cite{Lazarus2011}. \ks{Reproduced with permission from Soft Matter 7, 2552–2559 (2011). Copyright 2011 Royal Society of Chemistry.} 
    (e) Radial crack \cite{pauchard1999influence} and 
  (f) Circular cracks \citet{pauchard1999influence}. \ks{Reproduced with permission from Phys. Rev. E. 59, 3737 (1999). Copyright 1999 American Physical Society.}} 
    \label{fig_crack_3}
\end{figure}
A plethora of patterns are known for the desiccating colloidal deposit; they are polygonal~\cite{groisman1994experimental}, linear~\cite{allain1995regular}, wavy~\cite{goehring2011wavy}, spiral~\cite{Lazarus2011}, radial~\cite{pauchard1999influence} and circular~\cite{pauchard1999influence} as shown in Fig.~\ref{fig_crack_3}(a)-(f). In most cases, the physical quantities, such as -- spacing between two cracks $(\lambda)$, gap between cracks $(\delta)$, thickness of cracking material $(h)$, length of the crack $(L)$, the size of the flaw, area of cracking domain ($A_d$) and desiccation time characterizes the emerging crack pattern. These parameters allow us to study quantitatively the physical attribute of the crack patterns. For example, \citet{komatsu1997pattern} and \citet{allain1995regular} reported that for parallel cracks in the colloidal deposit $\lambda \sim h^{2/3}$, while \citet{goehring2013plasticity} reported $\delta \sim L^{1/2}$. 
\begin{figure}[ht]
    \centering
    \includegraphics[width=\linewidth]{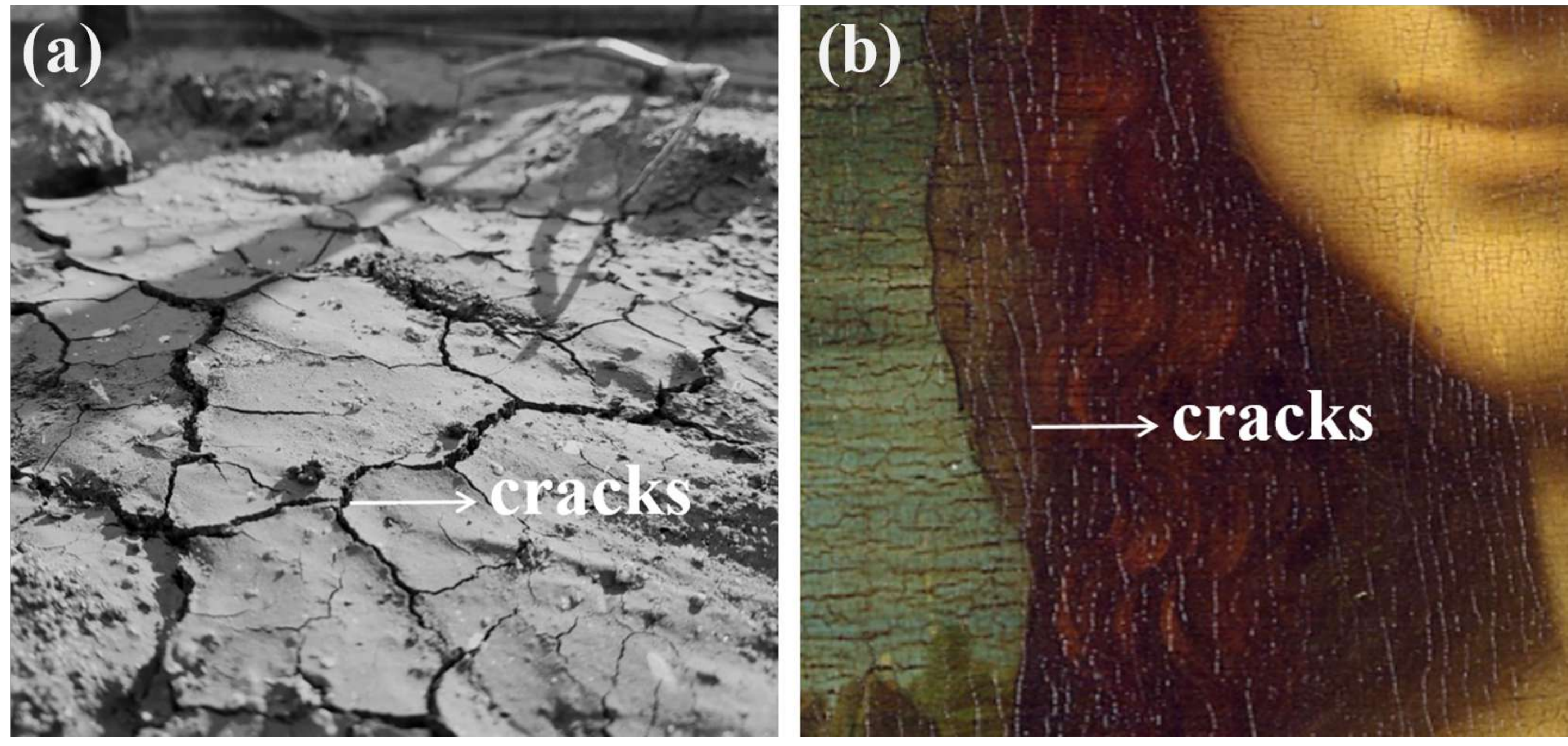}
    \caption{The picture of (a) dried mud with cracks (taken at IIT Madras) and (b) the celebrated painting exhibiting linear cracks are shown [Taken from:htttp:www.fast.u-psud.fr/pauchard/]. }
    \label{fig_crack_4}
\end{figure}
Interestingly, similar types of the emergence of cracks into distinct patterns are also common in nature, e.g. polygonal cracks in dried clay,~\cite{groisman1994experimental, goehring2009nonequilibrium, ma2019universal} interconnected cracks in old paintings~\cite{giorgiutti2016painting} as shown in Fig.~\ref{fig_crack_4}. The analogous properties of desiccation cracks exhibited by the colloidal deposit and the nature have urged researchers to investigate for a universality theory relating them \cite{goehring2009nonequilibrium}. \ks{The emergence of cracks into various distinct patterns is affected by numerous parameters. A few of them are physicochemical condition of drying,~\cite{giorgiutti2014elapsed,piroird2016role}, mechanical property of colloidal deposit,~\cite{giorgiutti2014elapsed,tirumkudulu2005cracking,karnail_singh2007} and local micro-structure constituting particles~\cite{piroird2016role,lama2018desiccation}}.
The role of these aforementioned parameters on the nucleation of crack and their manifestation into different patterns are described in the subsequent section. 

\subsubsection{Physicochemical condition} 
Various physicochemical parameters namely salinity~\cite{pauchard1999influence} and pH of colloidal suspension~\cite{Haque_2020}, relative humidity~\cite{giorgiutti2014elapsed} and temperature~\cite{lee2006temperature,lama2017tailoring} are reported to affect the desiccation crack pattern. The crack pattern is distinct for a specific physicochemical condition. These physicochemical parameters directly affect the drying kinetics, due to which the colloidal particle consolidation process is either accelerated or decelerated. This further influences the magnitude of strain energy that develops in the dried deposit. 
\begin{figure}[ht]
    \centering
    \includegraphics[width=\linewidth]{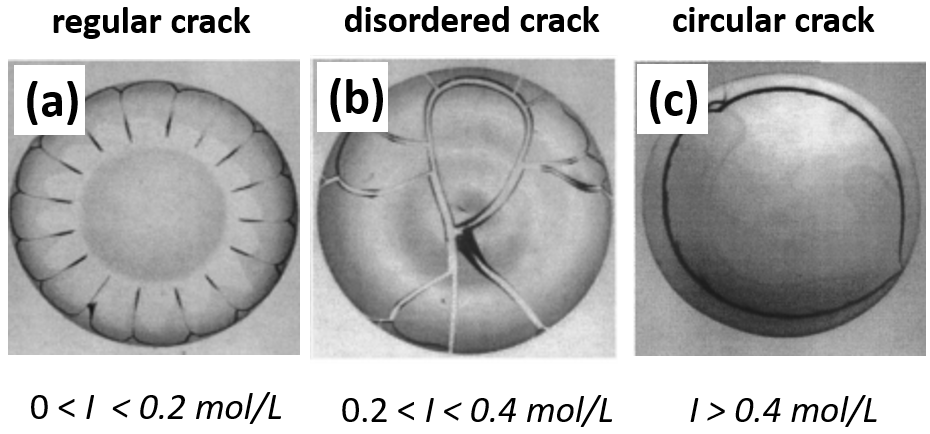}
    \caption{ Optical microscopy images of dried sessile drops comprising of silica particles (diameter $\sim 15$\,nm, initial volume fraction $\sim 0.2$) with different ionic strength \textit{I}\,(mol/L) and evaporating in an ambient environment. The dried deposit (a) with $0<I<0.2 \rm \,mol/L$ exhibits regular radial crack, (b) $0.2 < I < 0.4 \rm \,mol/L$ show disordered crack and (d) for $I>0.4 \rm \,mol/L$ cracks are circular. The drop base diameter is around 3\,mm \cite{pauchard1999influence}. \ks{Reproduced with permission from Phys. Rev. E. 59, 3737 (1999). Copyright 1999 American Physical Society.}} 
    \label{fig_crack_5}
\end{figure}
The characteristic time scales associated with the desiccation process typically characterize the dependence of physicochemical aspects on desiccation cracks. They are (i) gelation time ($t_G$), which is the time required for a drying dispersion to transform into a semi-solid and (ii) desiccation time ($t_D$), which corresponds to the time required for dispersion to transform into a solid completely. For example, \citet{pauchard1999influence} have demonstrated for a colloid drying in a sessile geometry, the addition of salt fastens the gelation time such that $t_G \ll t_D$, with all other drying conditions remaining the same (shown in Fig.~\ref{fig_crack_5}). This eventually varies the resultant morphology of crack. Interestingly, with a varying ionic strength in the dispersion, the crack pattern exhibited by the deposit are distinct (radial or disordered or circular) as shown in Fig.~\ref{fig_crack_5}. Besides, increasing or decreasing the relative humidity of the environment can analogously manipulate the  $t_G$. \citet{giorgiutti2014elapsed} has reported that $t_G \ll t_D$ for the colloid drying in sessile drop configuration. Furthermore, they show that increasing the RH of the drying environment results in the suppression of cracks, as shown in Fig.~\ref{fig_crack_6}. \ks{In addition to the abovementioned factors, the increasing desiccation temperature has been reported to decrease the desiccation time $t_d$ and alters the resultant crack pattern.~\cite{lama2017tailoring}}
\begin{figure}[ht]
    \centering
    \includegraphics[width=\linewidth]{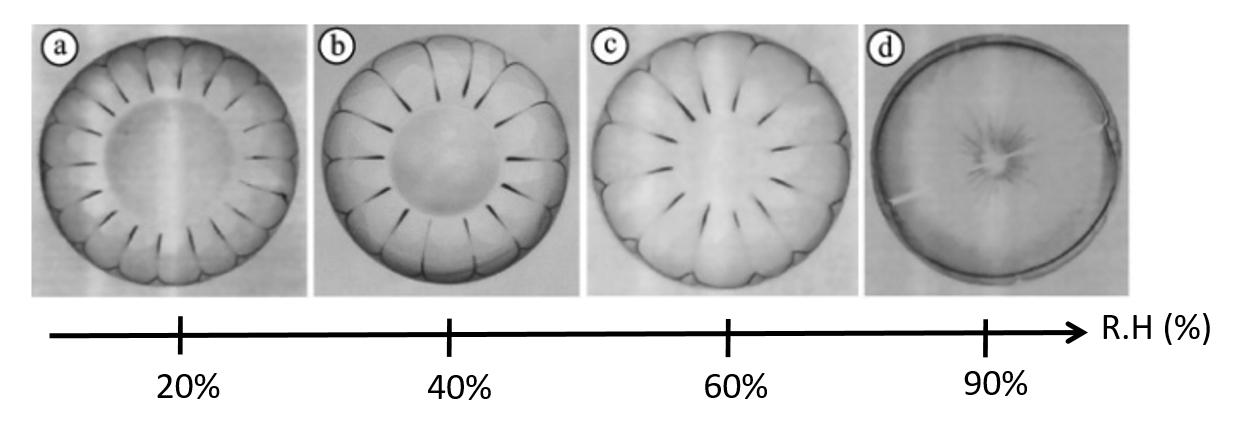}
    \caption{Optical microscopy images of the dried drops comprising of colloids (Ludox$^\copyright$ HS-40, $\theta_e = 40^\circ$) showing cracks. These deposits are for the droplets that are dried at room temperature but with RH (a) 20\%, (b) 40\%, (c) 60\% and (d) 90\%. The number of cracks can be seen to decrease with the increase in RH. The base diameter of each drop is around 3\,mm \cite{giorgiutti2014elapsed}. \ks{Reproduced with permission from Eur. Phys. J. E 37, 1–7 (2014). Copyright 2014 Springer.}} 
    \label{fig_crack_6}
\end{figure}
\subsubsection{Mechanical property of the colloidal deposit} 
Since the critical strain energy release rate ($G_c$) depends on the material properties such as $E_f, \nu_f, K$ of the deposit, the nucleation of cracks also depends on these quantities. In principle, the magnitude of these quantities, such as $E_f$ $\nu_f$, $K$, in a desiccating colloidal deposit, are always modulated, but towards the tail end of drying, i.e. when the crack nucleates, they are nearly constant. \citet{tirumkudulu2005cracking,karnail_singh2007} have derived an expression of critical stress $\sigma_c$ for a thin colloidal deposit drying isotropically. It is expressed as \cite{tirumkudulu2005cracking,karnail_singh2007}
\begin{equation}
    \frac{\sigma_c R_p}{2\gamma} = 0.1877 \Big( \frac{2R_p}{h}\Big)^{2/3} \Big(\frac{\mathcal{G}(E_f \nu_f) M\phi_{rcp} R_p}{2\gamma}\Big)^{1/3},
    \label{eq2_11}
\end{equation}
where, $R_p$ denotes radius of constituting particles, $\gamma$ is the surface tension of air-water interface, $M$ is coordination number, $\mathcal{G}(E_p, \nu_p)$ represents the shear modulus and $h$ denotes the thickness of the colloidal deposit. Note that the critical stress expressed in Eq.~(\ref{eq2_11}) was derived from the \textit{Routh-Russel} model. Here, we can obtain the corresponding strain from the critical stress in Eq.~(\ref{eq2_11}). In experiments, the mechanical property of colloidal deposits is manipulated either by changing the constituents or by varying the random close packing fraction ($\phi_{rcp}$) of the particles or by altering the drying rate ($V_E$).
\begin{figure}[ht]
    \centering
    \includegraphics[width = \linewidth]{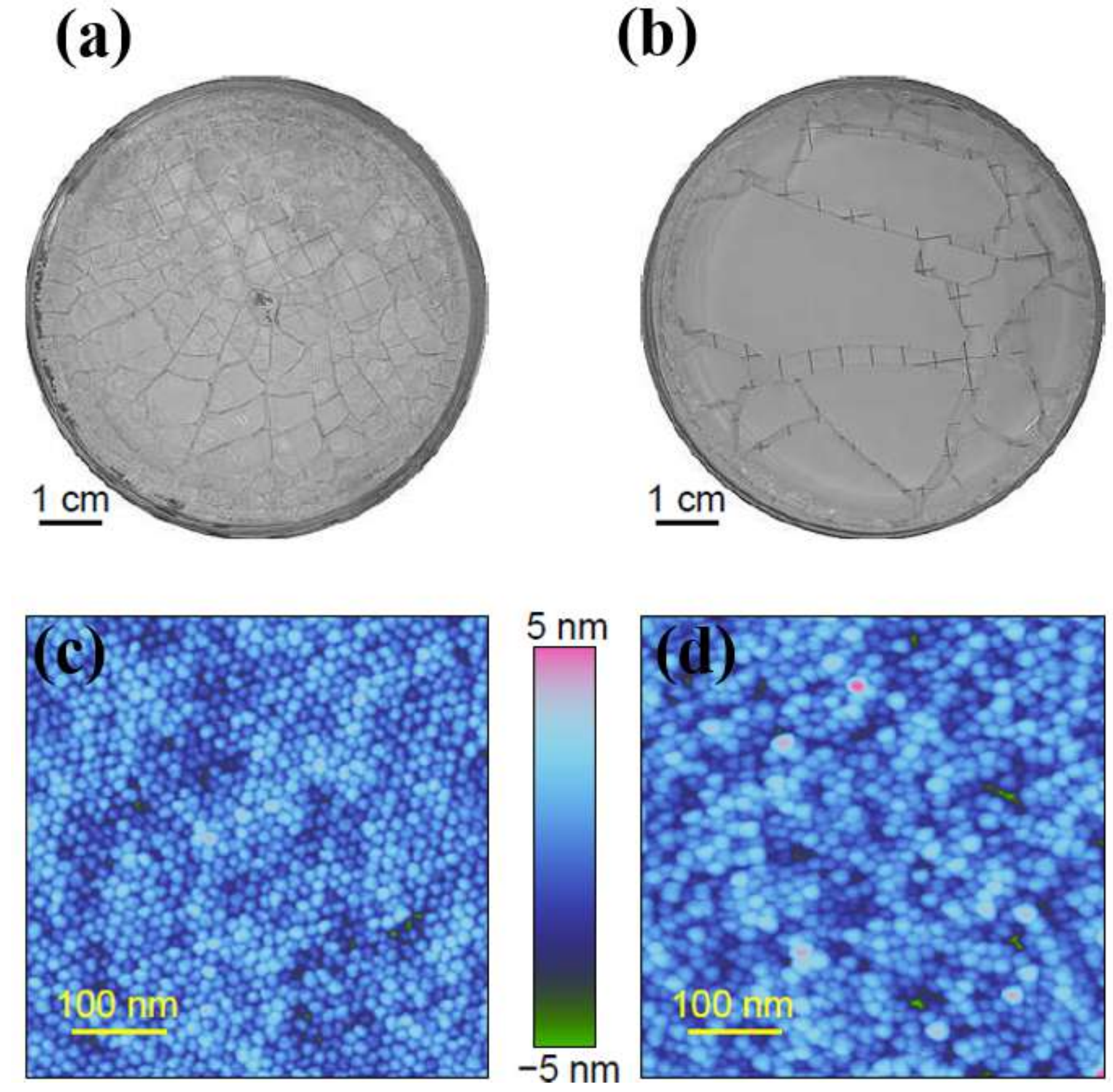}
    \caption{The picture of dried deposits on a petri dish with cracks. The deposit comprises Ludox$^\copyright$ HS-40 with an initial volume fraction of $\sim 0.4$. The picture of the dried deposit with cracks obtained by drying (a) at RH\,$\sim$\,10\% and (b)  RH\,$\sim$\,95\%. (c), (d) Atomic force microscopy (AFM) images for the respective deposits depict their local micro-structure \cite{piroird2016role}.}  
    \label{fig_crack_7}
\end{figure}

\subsubsection{Microstructure} 
The particle microstructure is an important parameter that dictates the crack pattern in the colloidal deposit. For example, \citet{piroird2016role} has reported the correlation between the number density of cracks (number of crack per unit length) and the local microstructure, i.e. the arrangement of the particles. Their observations suggest that the crack density is higher when the particles are hexagonally close-packed and it is significantly low when the constituting particles were randomly arranged in the dried 
 deposit (Fig.~\ref{fig_crack_7}(a)-(b)). They modulate the particle arrangement by varying the drying rate $V_E$. Drying at a higher rate resulted in the ordered assembly of closely packed particles (Fig.~\ref{fig_crack_7}(c)), while the slower drying rate yields an amorphous structure (Fig.~\ref{fig_crack_7}(d)). Note that in their experiments, the shape of constituting particles were spherical. 
\begin{figure}[ht]
    \centering
    \includegraphics[width=\linewidth]{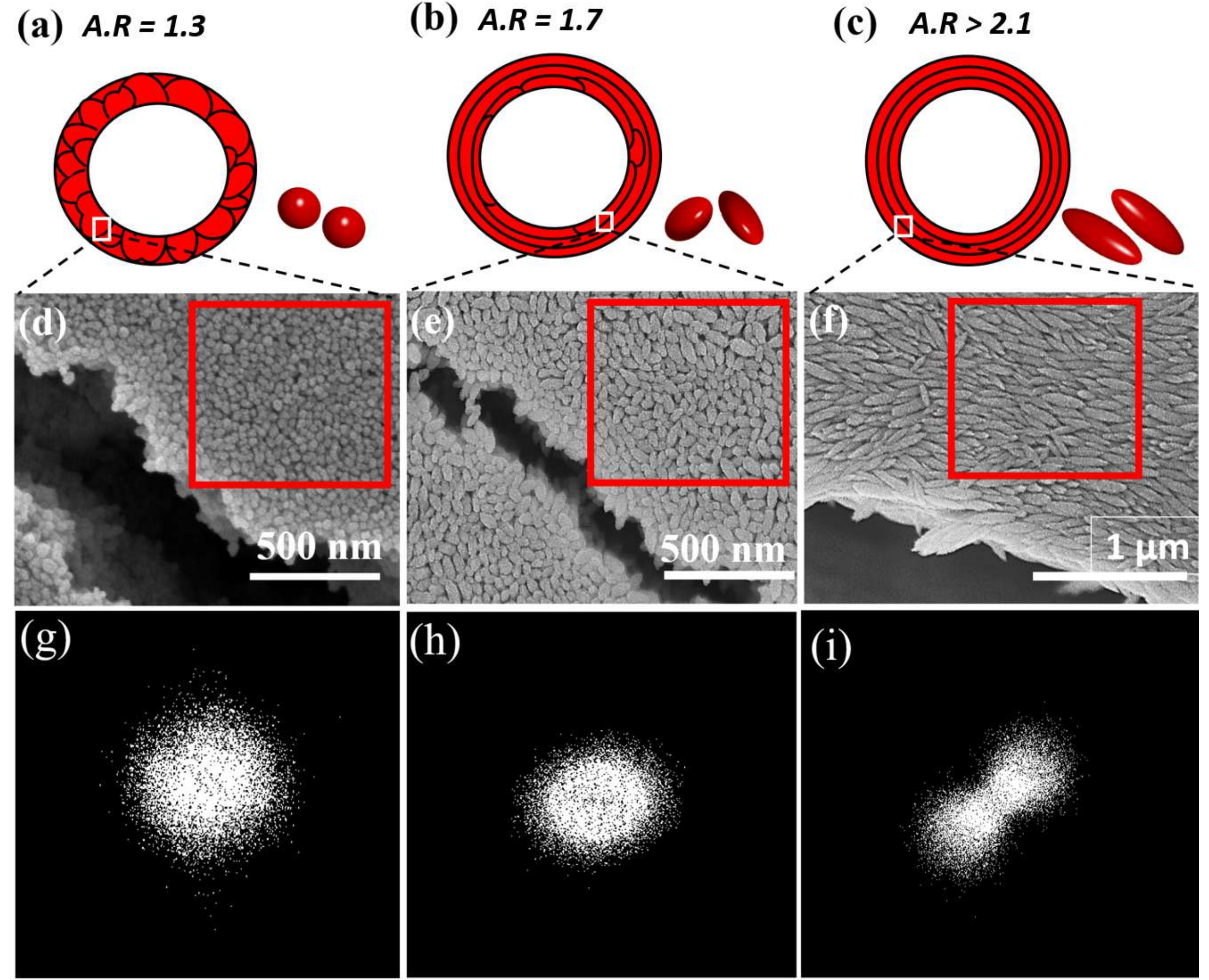}
    \caption{(a)- (c) Schematic of the dried ring-like deposit obtained by drying colloids (hematite ellipsoids, pH\,$\approx2$) in sessile drop geometry at $T_s \approx 25^\circ$C and RH\,$\approx$\,40\%. Scanning electron microscopy (SEM) images showing the region in the vicinity of cracks for the deposit comprising of particles with  SEM micrographs depict the particle assembly in the vicinity of the cracks for the deposits with particle aspect ratio with (d) AR\,$\approx$\,1.3, (e)AR\,$\approx$\,1.7, and (f) AR\,$>$\,1.3. (g)–(i): The fast Fourier transform (FFT) pattern of the deposit surface in the region marked by the rectangle in the respective SEM micrographs. The clear transition in the micro-structure from an isotropic phase to the anisotropic phase resulting in the variation of crack morphology from radial to circular is observed \cite{lama2018desiccation}. \ks{Reproduced with permission from Phys. Rev. Mater. 2, 085602 (2018). Copyright 2018 American Physical Society.}} 
    \label{fig_crack_8}
\end{figure}

For a deposit with non-spherical particles, the local arrangement of particles can vary between an isotropic phase and the nematic phase~\cite{lama2018desiccation}. We can characterize the degree of disorderedness in the particle arrangement by quantifying the 2D orientation order parameter $S$, which can be expressed as~\cite{nandakishore2016crack}
\begin{equation}
 S = <\cos{2\psi}>,   
\end{equation}
where $\psi$ is an angle between the major axis of a non-spherical particle and the reference axis. Here, the value of $S$ varies between $0-1$. For perfectly isotropic assembly of particles, $S = 0$, while for the nematic-like phase $0.4<S<0.9$ and $S = 1$ for perfect crystals. In the drying deposit of non-spherical particles, varying the aspect ratio $(AR)$ of the constituting particle alters the microstructural order. \citet{lama2018desiccation} has reported that the deposit with particles of $AR  > 2.1$ exhibits nematic-like ordering while that with $AR <1.7$, particles are randomly arranged. The resultant crack morphology was found to be significantly different as shown in Fig.~\ref{fig_crack_8}(a)-(i)). The deposit with particles of $AR > 2.1$ exhibits a circular crack, while the deposit with $AR < 1.7$ shows radial or interconnected cracks. 
   
\subsubsection{Cracks under external field} 

\begin{figure}[htb]
    \centering
    \includegraphics[width=\linewidth]{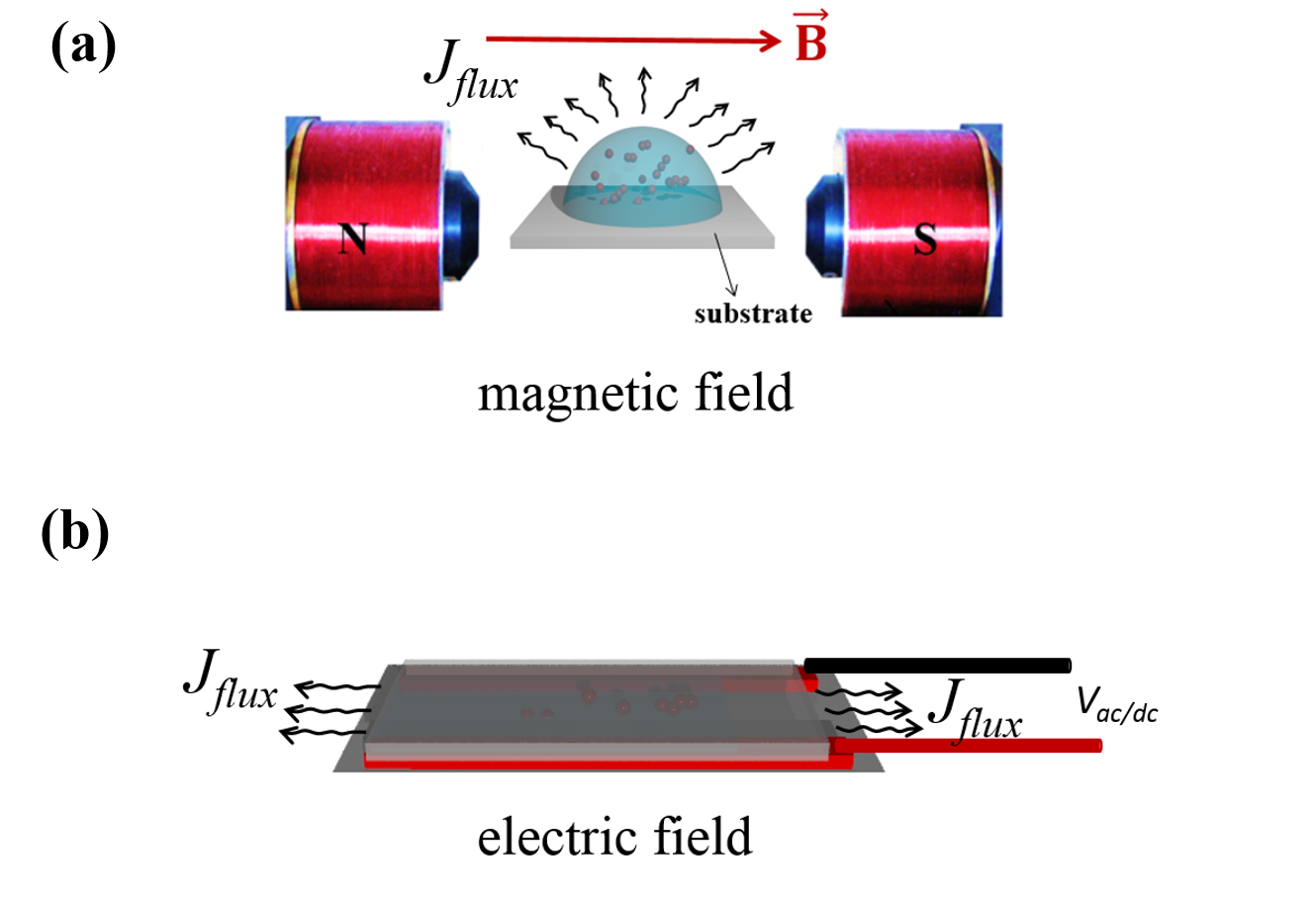}
    \caption{Schematic of standard drying geometry with an external field - depicts the drying of colloidal disperison (a) under an in-plane dc magnetic field $\Vec{B}$ and (b) under an electric field $\Vec{E}$. The images are taken from the PhD thesis of Hisay Lama, IIT Madras.}
    \label{fig_crack_9}
\end{figure}

Desiccation cracks under an external field are one of the modular routes to generate long-range spatial periodicity in cracks. The field-driven rearrangement of particles in the deposits can obtain the periodicity in cracks. The external field typically forces the particles to alter their orientation along the direction of the external field. The selection of an external field to drive these particles requires them to respond to the field. In the laboratory, magnetic field \cite{ding2009fabrication,martinez2016orientational,ngo2008cracks} and the electric field \cite{mittal2009electric,khatun2012electric,kumar2017stress} are prevalent options to drive the particles in the dispersion. The typical experimental set-up adopted for desiccation under magnetic and electric fields are depicted in Fig.~\ref{fig_crack_9}(a)-(b). 

\begin{figure}[htb]
    \centering
    \includegraphics[width=\linewidth]{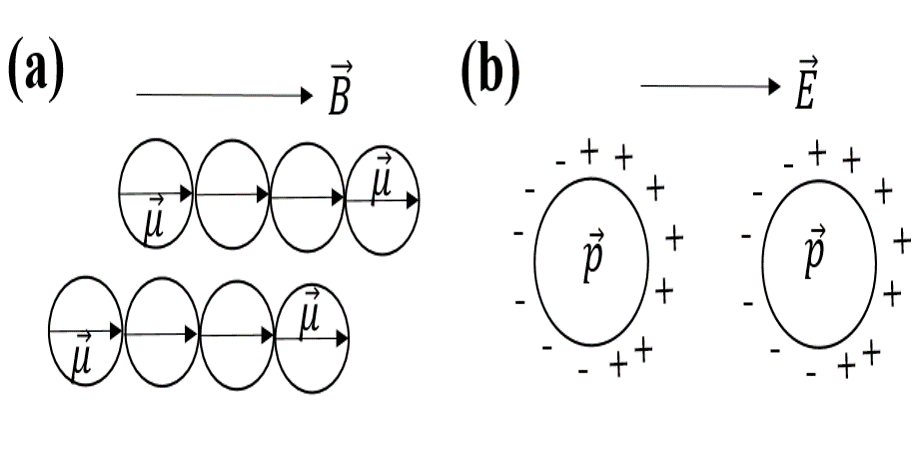}
    \caption{Schematic shows the external field-driven alignment of spherical particles. (a) Alignment of a magnetic particle with an intrinsic magnetic moment $\vec{\mu}$ under a magnetic field $\vec{B}$. (b) Charged particles under an electric field $\vec{E}$ that induces dipole moment $\vec{p}$. These images are borrowed from PhD thesis of Hisay Lama, IIT Madras.}
    \label{fig_crack_10}
\end{figure}

The colloid comprising magnetic particles under the uni-axial DC magnetic field ($\vec{B}$) tends to reorient them to form a chain-like structure. Customarily, the cracks in a deposit comprising spherical magnetic particles align along the direction of the applied field. For a deposit comprising of maghemite ($\gamma -Fe_2 O_3$) or hematite ($\alpha-Fe_2O_3$) particles, under a uniaxial DC field is reported to exhibit cracks that are parallel to the direction of $\vec{B}$~\cite{ngo2008cracks}. The magnetic particles in the dispersion typically possess the intrinsic magnetic moment $\vec{\mu}$ and interact with each other. Upon an application of dc magnetic field $\vec{B}$, the particles with an intrinsic magnetic moment $\vec{\mu}$ reorient themselves such that $\vec{\mu} \,||^l\,\vec{B}$, as shown in Fig.~\ref{fig_crack_10}(a). The net magnetic field-driven energy experienced by the particles is given by
\begin{equation}
    U_M = \Sigma_s \vec{\mu}_s . \vec{B} + U^M _{int},
\end{equation}
where $s$ is the index counting the particle and $U^M _{int}$ is the dipolar interaction energy. When $\vec{B}$ is sufficiently high, such that $U_M$ is greater than the thermal energy and hydrodynamic-driven effects, the particles exhibit magnetic field-driven alignment followed by the change in orientation of cracks.

Similarly, the desiccation under the electric field is another popular technique that generates spatially regular cracks~\cite{crassous2014field,trau1997assembly}. The cracks align along the direction of electric lines of force. Applying an electric field in the dispersion comprising charged particles typically manipulates their arrangement. These particles interact with each other via electrostatic interactions. When an electric field, $\vec{E}$, is applied to the dispersion, constituting charged particles get electrically polarized, and the charge distribution on the surface becomes asymmetrical with the particles possessing dipole moment $\vec{p}$. The net energy associated with the dispersion is
\begin{equation}
    U_E = \Sigma_s \vec{p}_s. \vec{E} + U^E _{int},
\end{equation}
where $U^E _{int}$ interaction energy between the dipoles. Analogous to the magnetic field. It is to be noted that when $U_E$ is greater than thermal and hydrodynamic-driven effects, particles and the resultant cracks undergo field-driven alignment.

\section{Summary}

In this review, we presented a brief overview of the physics of drying, primarily focusing on the drying of pure and binary liquid droplets. The consolidation phenomenon during the drying of colloidal drops, the formation of patterns in the resulting dried deposit, and the nucleation of desiccation cracks are discussed. The topics covered in this review are relevant for understanding physical phenomena like self-assembly, pattern formation, and desiccation cracks. The drying of colloids involves transforming the physical state from a liquid to a solid deposit. This is related to the evaporation-induced vapour diffusion into the surrounding atmosphere and the accumulation of particles owing to the local flow fields. The present study illustrated the flow fields in a sessile drop configuration. The formation of dried patterns such as coffee rings, uniform deposits, and multiple rings and their correlation to various physical parameters are discussed. Finally, recent research that emphasizes the impact of geometrical confinement on the patterning of colloids is highlighted.

Additionally, we demonstrated the drying-induced desiccation of cracks. In the early stages of the drying process, the consolidation of a drying dispersion into a solid deposit generates tensile stress, which leads to deformation and an accumulation of strain energy. The release of the excess strain energy results in the nucleation of cracks. The morphology and organization of these cracks depend on various parameters, such as physicochemical drying conditions, the mechanical properties of the particulate deposit, and then the local microstructural rearrangements. Finally, we explore the influence of external fields in controlling the formation of ordered cracks in the deposit. Tailoring the resulting dried particle deposit and cracks to create a specific pattern has potential applications in material development, including photonic crystals and lithographic templates. To achieve this, it is crucial to have a thorough understanding of the theoretical and experimental aspects of drying-driven phenomena. Therefore, this review provides an introductory background and insight into these subjects and will help understand several practical applications, namely, biological systems, medicine and forensics applications, combustion, ink-jet printing, hot-spot cooling, droplet-based microfluidics and coating technology.





\section*{Acknowledgements:} K.C.S. thanks the Science \& Engineering Research Board, India for the financial support through grant CRG/2020/000507. H.L. acknowledges JSPS Kakenhi Startup Grant (grant no. 21K20350).

\vspace{-0.5cm}

\section*{Data Availability Statement}
The data that support the findings of this study are available from the corresponding author upon reasonable request.

\section*{REFERENCES}
\nocite{*}


%

\end{document}